\documentclass[fleqn,usenatbib]{mnras}
\PassOptionsToPackage{pdfpagelabels=false}{hyperref} 
\pdfoutput=1
\usepackage{graphicx,color}
\usepackage{epstopdf}
\usepackage[latin1]{inputenc}
\usepackage[T1]{fontenc}
\usepackage{ae,aecompl}
\usepackage{graphics}
\usepackage{amsfonts}
\usepackage{amsmath}
\usepackage{multicol}
\usepackage{layout}
\usepackage{amssymb}
\usepackage[english]{babel}
\usepackage{scalerel}
\usepackage{times} 
\newcommand\reallywidehat[1]{\arraycolsep=0pt\relax
\begin{array}{c}
\stretchto{
  \scaleto{
    \scalerel*[\widthof{\ensuremath{#1}}]{\kern-.5pt\bigwedge\kern-.5pt}
    {\rule[-\textheight/2]{1ex}{\textheight}} 
  }{\textheight} %
}{0.5ex}\\           
#1\\                 
\rule{-1ex}{0ex}
\end{array}
}
\title[Fast Weak Lensing Simulations]{Fast Weak Lensing Simulations with Halo Model}
\author[Giocoli C. et al. 2017]{\parbox{\textwidth}{
    Carlo Giocoli$^{1,2,3,4}$\thanks{E-mail:\href{mailto:carlo.giocoli@unibo.it}
    {carlo.giocoli@unibo.it}}, Sandra Di Meo$^5$, Massimo Meneghetti$^{3,4}$,
    Eric Jullo$^2$, Sylvain de la Torre$^2$, Lauro Moscardini$^{1,3,4}$, Marco Baldi$^{1,3,4}$,
    Pasquale Mazzotta$^5$, R. Benton Metcalf$^{1}$
  }\\ \\
$^1$Dipartimento di Fisica e Astronomia, Alma Mater Studiorum Universit\`{a} di 
  Bologna, via Gobetti 93/2, 40129, Bologna, Italy \\
$^2$Aix Marseille  Univ,  CNRS,  LAM, Laboratoire  d'Astrophysique  de
Marseille, Marseille,  France \\
$^3$INAF - Osservatorio Astronomico di Bologna, via Ranzani 1, 40127, Bologna, Italy \\ 
$^4$INFN - Sezione di Bologna, viale Berti Pichat 6/2, 40127, Bologna, Italy \\
$^5$Dipartimento di Fisica, Universit\'{a} degli Studi di Roma ``Tor Vergata'',
via della Ricerca Scientifica 1, 00133 Roma, Italy
}
\pubyear{2016}

\begin{document}
\label{firstpage}
\pagerange{\pageref{firstpage}--\pageref{lastpage}}
\maketitle

\begin{abstract}
Full ray-tracing maps of gravitational lensing, constructed from
N-Body simulations, represent a fundamental tool to interpret present
and future weak lensing data.  However the limitation of computational
resources and storage capabilities severely restrict the number of
realizations that can be performed in order to accurately sample both
the cosmic shear models and covariance matrices.  In this paper we
present a halo model formalism for weak gravitational lensing that
alleviates these issues by producing weak-lensing mocks at a
reduced computational cost. Our model takes as input the halo population
within a desired light-cone and the linear power spectrum of the
underlined cosmological model. We examine the contribution given by
the presence of substructures within haloes to the cosmic shear power
spectrum and quantify it to the percent level.  Our method allows us to
reconstruct high-resolution convergence maps, for any desired source
redshifts, of light-cones that realistically trace the matter density
distribution in the universe, account for masked area and sample
selections. We compare our analysis on the same large scale structures
constructed using ray-tracing techniques and find very good agreements
both in the linear and non-linear regimes up to few percent levels.
The accuracy and speed of our method demonstrate the potential of our
halo model for weak lensing statistics and the possibility to generate
a large sample of convergence maps for different cosmological models
as needed for the analysis of large galaxy redshift surveys.

\end{abstract}

\section{Introduction}

Cosmological surveys - e.g. VVDS, COSMOS, VIPERS, BOSS, DES
\citep{des,sousbie08,sousbie11,guzzo14,percival14,lefevre15,codis15}
- and observations from long-term space missions such as the HST
telescope, Chandra and XMM are delivering to the scientific community
a very large quantity of data which seem to be quite well interpreted
by a standard cosmological model in which two unknown forms of matter
and energy - named dark matter and dark energy - dominate the energy
content of our Universe.  However the analyses recently performed by
the KiDS collaboration on the KiDS-450 dataset \citep{hildebrandt17}
have reached results in good agreement with other low redshift probes
of large scale structure \citep[for example the CFHTLenS data analyses
  presented by][]{benjamin13,heymans13,hildebrandt12,
  kilbinger13,kitching14} and pre-Planck CMB measurements -- like ACT,
SPT and WMAP9 \citep{wmap9} -- confirming the tension with the 2015
Planck outcomes \citep{planck16a}. It is interesting to point out that
if the tension between those cosmological probes persists in the
future modification of the current concordance model will become
necessary. 

The inhomogeneities and redshift evolution of non-linear structures in
the universe can be evaluated using the statistical measurements of
the ellipticity of background galaxies. The determination of the
galaxy shapes and redshifts, in the absence of systematic errors, can
be translated into an unbiased measurement of the shear
\citep{melchior11,bartelmann12}, which can be used to reconstruct the
projected matter density distribution along the line of sight
\citep{kaiser93,kaiser95,viola11}. Tomographic reconstruction of the
matter density field and their statistical properties can be then
employed to constrain standard cosmological parameters \citep[as e.g
  the matter density parameter $\Omega_{\rm m}$ and the initial power
  spectrum normalization $\sigma_8$,
  see][]{fu08,kilbinger13,hildebrandt17} as well as possible
parameterizations of the dark energy equation of state
\citep{kitching14,kitching15,kohlinger15}.

For this reason, cosmic shear measurements from weak gravitational
lensing effect represent a primary probe for many ongoing and future
wide field surveys \citep{des,flaugher05,wfirst,ivezic08,ivezic09}
and in particular for the wide field survey covering 15,000
sq. degrees that will be performed by Euclid \citep{euclidredbook}.
In this context, it is very important to have the possibility to
construct flexible reference models of weak lensing statistics that
can account for finite survey areas, masking and sample selection, as
well as probe high redshift regimes. In particular it is imperative to
be able to perform a large sample of independent simulations of weak
lensing statistics for the need of well sample the covariance matrix
to keep systematics and possible biases that may appear in the
measurements under control.  Cosmological numerical simulations of
large scale structures, from which we can reconstruct realistic past
light-cones up to a desired source redshift, represent the natural
reference tools to build weak lensing models
\citep{jain00,vale03,sato09,hilbert09}. They give the possibility not
only to correctly model the structure formation processes as a
function of the cosmic time but also to include self-consistent
recipes to model the baryonic physics: cooling, star formation
activities and the various types of feedback processes
\citep{hirschmann14,beck16}. Numerical simulations also allow for
exploration of a large variety of cosmological parameter spaces as
well as to model the structure formation mechanisms in non-standard
cosmological scenarios.  Nonetheless, all these interesting phenomena
that can be studied with numerical simulations require tuning the
numerical setup in order to find the best compromise between the size
of the numerical simulation box and number of snapshots saved -- which
set the maximum redshift up to which a statistically unbiased light
cone can be constructed and the largest modes of the density field
that can be probed -- and the particle mass which defines the
resolution for the modeling of small scale signals. Typical analyses
performed thus far properly model the statistical properties of the
weak lensing field up (down) to modes $l\approx 10^4$ (arcminute
scales).

Recently \citet{giocoli16a}, within the BigMultiDark collaboration,
have created lensing maps up to redshift $z_s=2.3$ for the two VIPERS
fields W1 and W4 and computed their associated weak lensing covariance
matrices for different source redshifts.  The resolution of the grid
on which particles have been placed and through which the light-rays
have been shot have been chosen to be equal to $6$ arcsec.  This small
scale limit of the simulations is mainly set by the mass and force
resolution of the BigMultiDark simulation \citep{prada16}, which
allows for trustworthy the lensing measurements only down to $\sim
1.5$ arcmin. Recently \citet{delatorre16} have used as reference the
lensing predictions from the BigMultiDark light-cones together with
the redshift-space distortions from the final VIPERS redshift survey
dataset and galaxy-galaxy lensing from CFHTLenS with the aim of
measuring the growth rate of structure.  The resolution of the
analysis performed by \citet{harnois-deraps12} -- where the authors
have accurately measured non-Gaussian covariance matrices and set the
stage for systematic studies of secondary effects -- is only slightly
higher.  In the latter work, a set of $185$ high-resolution N-body
simulations was performed, and the corresponding past light-cones were
constructed through a ray tracing algorithm using the Born
approximation. In a subsequent work, \citet{harnois-deraps15b} -- and
also \citet{angulo15} -- have investigated the importance of finite
support -- related to the limited box size of the simulation and
possible small field of view when constructing the lensing light-cones
-- which may suppress the two-point weak lensing statistic on large
scales. However such issues may be circumvented by performing lensing
simulations consistently with the limited size and geometry of the
observed lensing survey, but including large scale modes using
approximated methods from linear theory
\citep{monaco13,tassev13,monaco16}. Recently also \citet{petri16a}
have shown that for weak lensing statistics the full ray tracing
simulation is indeed unnecessary and that simply projecting the
lensing planes causes negligible errors compared to this; in
particular \citet{petri16b} have re-cycled a single N-body box as many
as 10,000 times generating statistically independent weak lensing maps
with sufficient accuracy.

Particularly interesting is also the possibility to perform weak
lensing simulations in a variety of different cosmological models. For
example in this case the availability of numerical simulations of
structure formation for those models is a fundamental starting
point. In this respect, we mention the analyses performed in
non-standard models with coupling between Dark Energy and Cold Dark
Matter by \citet{giocoli15} and \citet{pace15}, that showed specific
signatures with respect to standard $\mathrm{\Lambda}$CDM mainly when
performing a tomographic weak lensing analyses. In the same direction
goes the work performed by \citet{tessore15} which have produced weak
lensing maps of large scale structure in modified gravity cosmologies
that exhibit gravitational screening in the non-linear regime of
structure formation. \citet{carbone16} have presented a
cross-correlation analyses of CMB and weak-lensing signals using
ray-tracing across the gravitational potential distribution provided
in massive neutrinos simulations. These authors find an excess of
power with respect to the massless run, due to free streaming
neutrinos, roughly at the transition scale between the linear and
non-linear regime.

The production of a large number of independent light-cones
realizations for different cosmological models is an essential tool
for the interpretation of the large wealth of weak lensing data, that
will become available in the next decades. It is also crucial to go
beyond the Gaussian assumption in the characterization of the weak
lensing error bars, both in the linear and non-linear regimes to
correctly assess the sensitivity of the weak lensing signal to
cosmological parameters.

In this context, it is important to stress that weak lensing
simulations have to be made consistent with the survey properties;
simulated light-cones in first analysis should mimic the geometry as
well as the masking of the survey area.  Usually many light-cone
realizations are needed in order to obtain a precise estimate of the
covariance matrices over a wide range of scales and for sources at
different redshifts, and all such realizations need to be extended to
the various cosmological models we would like to sample. A
comprehensive program of weak lensing analyses performed based on full
N-body simulations then requires enormous computational resources and
huge storage capabilities, which are difficult to access even at the
largest computing centers.

On the other hand, approximate methods are much faster, and less
memory demanding, hence opening the possibility to test various
cosmological scenarios at a highly reduced computational cost. In this
regard, it is interesting to mention the work by \citet{yu16} who have
presented a fast method to generate weak lensing maps based on the
assumption that a lensing convergence field can be Gaussianized to
excellent accuracy by a local transformation. Even if their
constructed maps have a good representation of the large scale
normalization of the cosmic shear power spectrum, they have larger
power at intermediate scales than the simulated reference fields and
vice versa at small scales.  These effects are probably due to the
imperfection of the Gaussian Copula Hypothesis on which their method
is based.

\begin{figure*}
  \includegraphics[height=7.25cm]{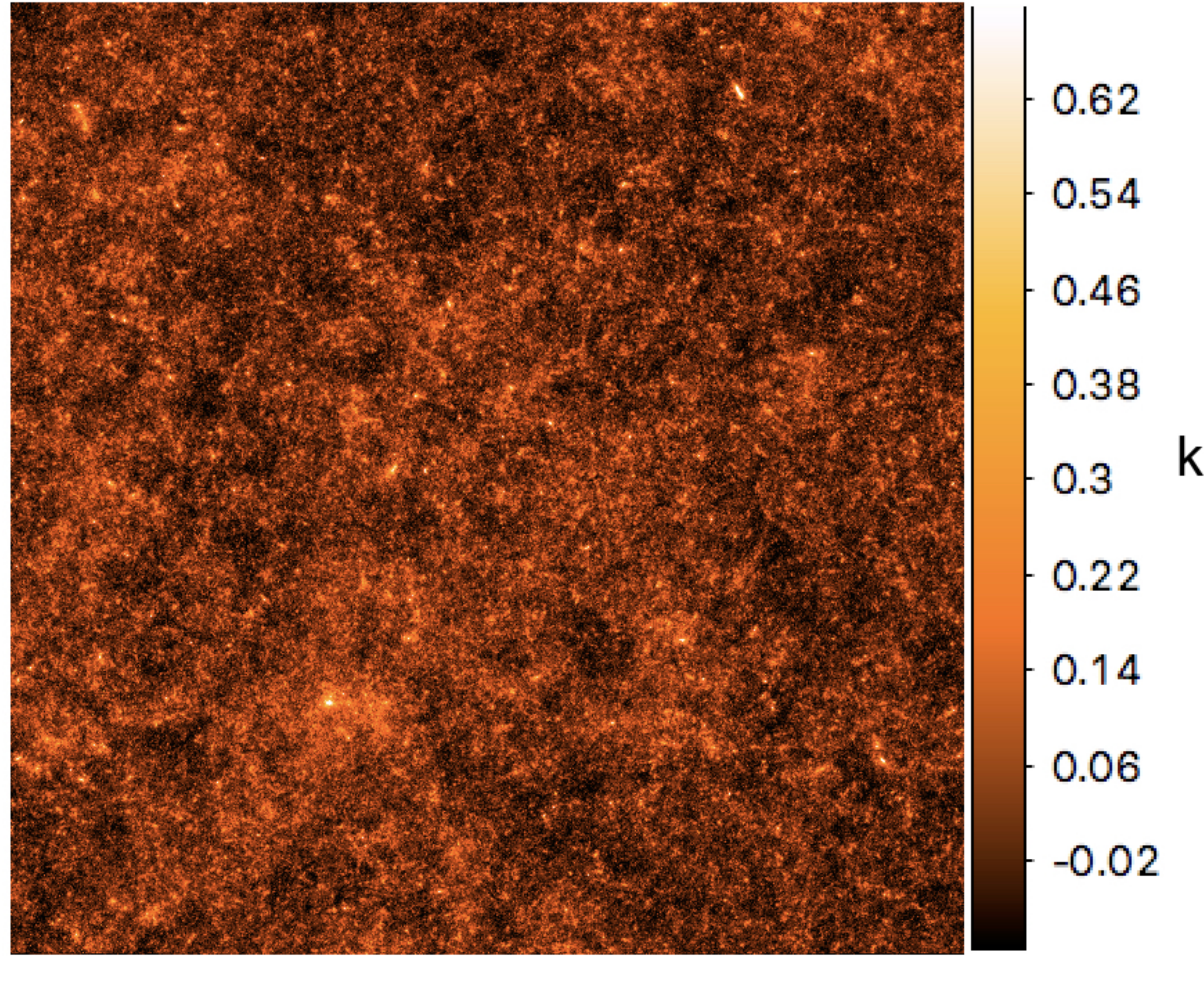}
  \includegraphics[height=7.25cm]{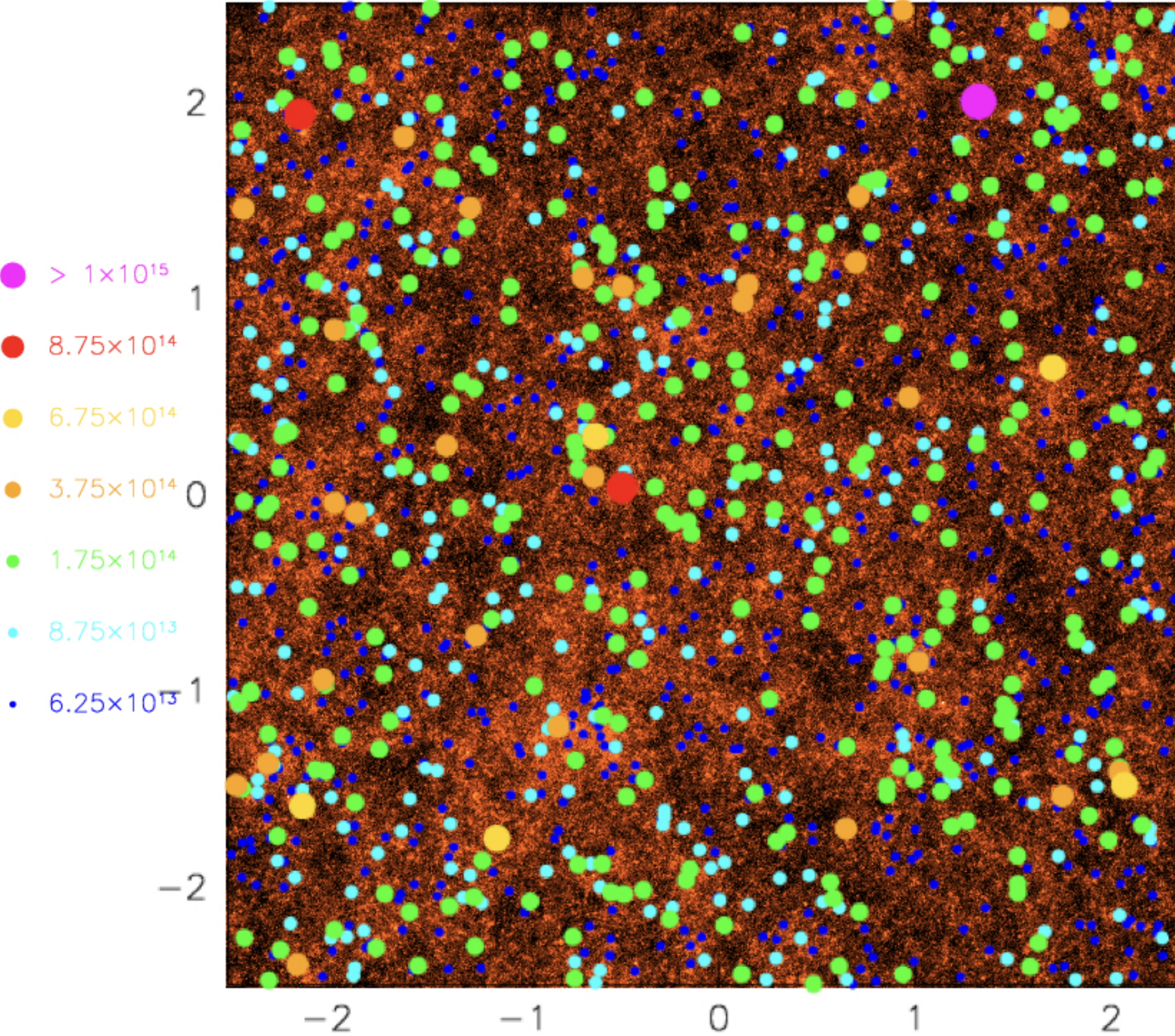}
  \caption{Left panel: convergence map ($5 \times 5$ sq. degrees) of a
    light-cone realization up to $z_s=4$ constructed using the
    multi-plane ray-tracing \textsc{glamer} pipeline.  Right panel:
    convergence map with over-plotted the haloes present within the
    light cone more massive than $5\times
    10^{13}\mathrm{M_{\odot}}/h$.  The various size coloured circles
    indicate haloes with different masses, as labelled in the plot.
    The masses refer to the FoF group definition.\label{figMapandFoF}}
\end{figure*}

Producing a large sample of realistic weak lensing simulations is
becoming a challenging but necessary task for interpreting the
outcomes of future wide field surveys.  Importantly those allow ($1$)
to mimic the survey geometry and masked regions ($2$) to consistently
sample the expected weak lensing signals from the matter density
distribution along the line of sight and ($3$) to construct reference
models using the observed source redshift distribution from a given
survey. A large number of light-cones plus weak lensing measurements
is needed to ensure a good sampling of the non-linear properties of
structure formation and to have under control the Gaussian and the
non-Gaussian terms and the cosmic variance in estimating the
covariance matrices \citep{harnois-deraps15,harnois-deraps15b}.

In this paper, we use the halo model formalism for weak gravitational
lensing, to quickly and accurately generate high-resolution
convergence maps for any desired field of view and source redshift
distribution in the context of a standard $\mathrm{\Lambda}$CDM
cosmological scenario. Similarly \citep{li02,giocoli12a,giocoli16b}
have used the lensing halo model formalism for strong lensing studies
while \citep{kainulainen11,lin15a,lin15b,matilla16} have used it for
weak lensing predictions.  The simulated maps can then be masked and
cut to reproduce the geometry of the observed survey. The weak lensing
statistical properties of the light-cones can also be sampled
according to a realistic source sample, their redshift distribution
and clustering. The extension of our method to a variety of
non-standard cosmological models will be investigated in a forthcoming
paper.

Our paper is organised as follows: in section~\ref{secmodel} we
present the reference numerical simulated light-cones with which we
compare our model and describe the idea of the method, in
section~\ref{sechmmodel} we present our halo model for weak
gravitational lensing and in section~\ref{secstatmoka} we define the
statistical estimators that we apply to our simulated light-cones to
characterise their properties.  In section~\ref{sumandcon}, we
summarise and discuss our results.

\section{Model}
\label{secmodel}

In this work we present a fast method to produce weak lensing
simulations using a halo model approach.  In our analysis we use the
halo catalogs corresponding to the particle light-cones extracted from
a reference cosmological simulation.  The light-cones have been
produced by remapping the simulated snapshots into cuboids and
projecting the particles into lens planes up to a given source
redshift. In this work we will make use of the halo and subhalo
catalogues to reconstruct the weak lensing field, using the halo
model, in a desired field of view and compare it with the prediction
obtained using the particles as tracers of the projected density. In
this way, we statistically reconstruct the matter density distribution
along the line-of-sight \citep{giocoli15,giocoli16a}, avoiding
replicating the same structures and producing gaps.  The convergence
maps have been computed from the projected lens planes using the
ray-tracing \textsc{glamer} pipeline \citep{metcalf14} as described in
\citet{petkova14}.

\subsection{The Numerical Simulation}

In this section we present the reference numerical simulation we adopt
and stress that our method is very general and ready to be applied to
any halo -- and subhalo -- catalogue.

The cosmological parameters of our reference simulation have been set
accordingly to the WMAP7 results. In particular, the numerical
simulation used here is the $\mathrm{\Lambda}$CDM run extracted from
the \textsc{CoDECS} suite \citep{baldi12b}, where the initial
conditions are generated using the \textsc{N-GenIC}
code\footnote{\url{http://www.mpa-garching.mpg.de/gadget}} by
displacing particles from a homogeneous '\emph{glass}' distribution in
order to set up a random-phase realisation of the linear matter power
spectrum of the cosmological model according to Zel'dovich
approximation \citep{zeldovich70}.  The particles displacements are
then rescaled to the desired amplitude of the density perturbation
field at some high redshift ($z_{i}=99$), when all perturbation modes
included in the simulation box are still evolving linearly.  This
redshift is then taken as the starting redshift of the simulation, and
the corresponding particle distribution as the initial conditions for
the N-body run. In setting the initial conditions for the simulation
we have chosen $\Omega_{\rm CDM} = 0.226$, $\Omega_b=0.0451$,
$\Omega_{\Lambda} = 0.729$, $h=0.703$ and $n_s=0.966$, the initial
amplitude of the power spectrum at CMB time ($z_{\rm CMB}\approx
1100$) $A_s(z_{\rm CMB}) = 2.42 \times 10^{-9}$ which correspond at
$z=0$ to $\sigma_8=0.809$.

The simulation has a box size of $1$ comoving Gpc/$h$ aside and
include $1024^3$ for both the components CDM and baryon for a total
particle number of approximately $ 2\times 10^9$. The mass resolution
is $m_{\rm CDM} = 5.84 \times 10^{10} \mathrm{M_{\odot}}/h$ for the
cold dark matter component and $m_b = 1.17 \times
10^{10}\mathrm{M_{\odot}}/h$ for baryons, while the gravitational
softening was set to $\epsilon_g = 20\;\mathrm{kpc}/h$.  Despite the
presence of baryonic particles this simulation does not include
hydrodynamics and is therefore a purely collisionless N-body run.

We stored about thirty snapshots between $z = 10$ and $z = 0$ at each
simulation snapshot, halos have been identified using
Friends-of-Friends (FoF) algorithm adopting a linking length parameter
$b = 0.2$ times the mean inter-particle separation of the CDM
particles as primary tracers of the local mass density, and then
attaching the baryonic particles to the FoF group of their nearest
neighbours.  Then, running \textsc{subfind} \citep{springel01b} -- on
each simulation snapshot, for each FoF-group we compute $M_{\rm 200}$
as the mass enclosing a sphere with density $200$ times the critical
density $\rho_c(z)$ at that redshift and assuming the particle with
the minimum gravitational potential as the halo
centre. \textsc{subfind} also searches for over-dense regions within a
FoF group using a local SPH (Smoothed Particle Hydrodynamics) density
estimate, identifying substructure candidates as regions bounded by an
isodensity surface that crosses a saddle point of the density
field. This algorithm is also testing that these possible
substructures are physically bounded with an iterative unbinding
procedure.  In what follows, we will indicate with $M_{\rm FoF}$ the
mass of the Friends-of-Friends group, with $M_{\rm 200}$ the mass of
the sphere enclosing $200$ times the critical density of the universe
and with $m_{\rm sub}$ the self-bound mass of substructures.

\subsection{Building the past-light-cone with \textsc{MapSim}}

\begin{figure*}
  \includegraphics[width=\hsize]{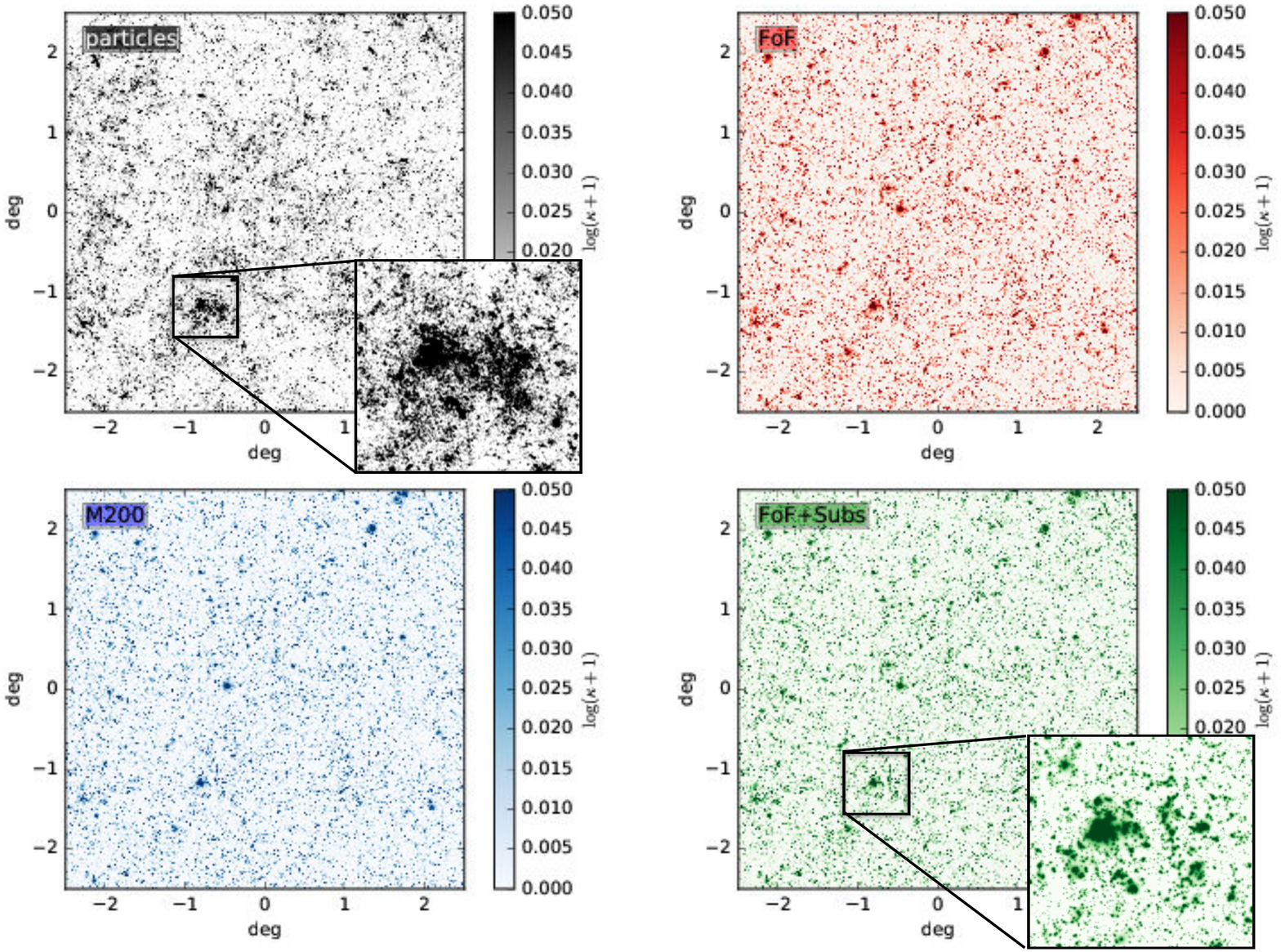}
  \caption{Convergence maps of a light-cone extending up to
    $z_s=4$. Top left panel: convergence maps created using
    ray-tracing in the light-cone constructed from the particles
    extracted from the simulation snapshots. Top right, bottom left
    and bottom right panels: convergence maps constructed using the
    halo model formalism based on $M_{\rm FoF}$, $M_{\rm 200}$ and
    $M_{\rm FoF}+m_{\rm sub}$ catalogs in the light-cone,
    respectively.\label{figmaps}}
\end{figure*}  

To build the lensing maps of the light-cone we piled together
different slices of the simulation snapshots up to $z_s=4$. The size
of the light-cone we consider has an angular aperture of $5$ deg,
which combined with the comoving size of the simulation box of $1$
Gpc/$h$, ensures to uniformly construct the mass density distribution
in redshift without gaps. For this purpose we use the \textsc{MapSim}
code \citep{giocoli15,tessore15} that extracts the particles from the
simulation's snapshot files and assembles them into a light-cone.  The
code initialises the memory and the grid size of the maps reading an
input parameter file. This file contains information about the desired
field of view (chosen to be $5$ deg on a side), the highest source
redshift (in this case $z_s = 4$) and the locations of the snapshot
files.  The number of required lens planes is decided ahead of time in
order to avoid gaps in the constructed light-cones and the available
stored simulation snapshots. We emphasize that in order to properly
statistically sample the evolution of the matter density distribution
as a function of the cosmic time within the light-cone we collapse in
each lens plane the closest snapshot in redshift.  The code, reading
each snapshot file at a time from low to high redshift, extracts only
the particle positions within the desired field of view and is not
much memory consuming since it needs to allocate only a single
snapshot file.  The lens planes are built by mapping the particle
positions to the nearest pre-determined plane, maintaining angular
positions, and then pixelising the surface density using the
triangular shaped cloud (TSC) method \citep{hockney88}. In
constructing the lens planes we try to preserve as much as possibile
the cosmological evolution of the structures by projecting into planes
the snapshot with the closest redshift.  The grid pixels are chosen to
have the same angular size on all planes, equals to $2048\times 2048$,
which allows to resolve approximately $8.8$ arcsec per pixel. The lens
planes have been constructed each time a piece of simulation is taken
from the stored particle snapshots; their number and recurrence depend
on the number of snapshots stored while running the simulation.  In
particular in running our simulation we have stored $17$ snapshots
from $z\sim4$ to $z=0$ reasonably enough to construct a complete
light-cone up to $z_s=4$ with $22$ lens planes.  The selection and the
randomisation of each snapshot is done as in \citet{roncarelli07} and
discussed in more details in \citet{giocoli15}.  If the light-cone
reaches the border of a simulation box before it reaches the redshift
limit where the next snapshot will be used, the box is re-randomised
and the light-cone extended through it again.  Once the lens planes
are created the lensing calculation itself is done using the
\textsc{glamer} pipeline \citep{metcalf14,petkova14}. Considering that
at low redshifts, where many massive haloes are present, we have saved
many snapshots -- for example we use twelve snapshots up to redshift
$z=1.2$ from which we produce fourteen lens planes -- when projecting
particles into separate lens planes we do not account for particle
clumps in haloes that are located on the slice boundaries with
particles on either side. As discussed by \citet{hilbert09} this
effect can eventually produce an over-counting of particles that may
bring a relative difference to the convergence power spectrum of
approximately $0.1\%$.

Defining ${\pmb \theta}$ the angular position on the sky and ${\pmb
  \beta}$ the position on the source plane (the unlensed position),
then a distortion matrix ${\bf A}$ can be defined as
\begin{align}                                                
{\bf A} \equiv \frac{\partial {\pmb \beta}}{\partial {\pmb \theta} } =         
\left(   
\begin{array}{cc} 
1-\kappa-\gamma_1 & \gamma_2  \\ 
\gamma_2  & 1-\kappa + \gamma_1 
\end{array}   
\right)\,,
\end{align} 
where $\kappa$ represents the convergence and the pseudo-vector ${\pmb
  \gamma}\equiv\gamma_1+i \gamma_2$ the shear.  In the case of a
single lens plane, the convergence can be written as:
\begin{equation}
\kappa({\pmb \theta}) \equiv \frac{\Sigma({\pmb \theta})}{\Sigma_{\rm
    crit}}\ , \label{eqconvergence}
\end{equation}
where $\Sigma({\pmb \theta})$ represents the surface mass density and
$\Sigma_{\rm crit}$ the critical surface density as:
\begin{equation}   
\Sigma_{\rm crit} \equiv \dfrac{c^2}{4 \pi G} \dfrac{D_l}{D_s D_{ls}},
\end{equation}                
where $c$ indicates the speed of light, $G$ the Newton's constant and
$D_l$, $D_s$ and $D_{ls}$ the angular diameter distances between
observer-lens, observer-source and source-lens, respectively.  In the
case of multiple lens planes the situation is slightly
different. After the deflection and shear maps on each plane are
calculated, the light rays are traced from the observers through the
lens planes up to the desired source redshift. The shear and
convergence are also propagated through the planes as detailed in
\citet{petkova14}.  \textsc{glamer} performs a complete ray-tracing
calculation that takes into account non-linear coupling terms between
the planes as well as correlations between the deflection and the
shear. However, for this work when running the ray-tracing pipeline we
have adopted the Born approximation, that is following the light-rays
along unperturbed paths.  As discussed in \citet{giocoli16a} -- by
performing a full ray-tracing comparison -- and in \citet{schaefer12}
-- by computing an analytic perturbative expansion -- the Born
approximation is an excellent approximation for weak cosmic lensing
down to very small scales ($l \geq 10^4$). We underline that the
physical modelling at these very small scales is far from the purpose
of this work and we are aware that it may eventually need a correct
and self-consistent treatment of the baryonic components
\citep{mohammed14,harnois-deraps15}. \\
   
In the left panel of Fig.~\ref{figMapandFoF} we show the convergence
map of the first light-cone realisation assuming a source redshift
$z_s=4$.  In order to have various statistical samples, we have
created $25$ light-cone realisations. They can be treated as
independent since do not contain the same structures along the
line-of-sight, considering the size of the simulation box $1$Gpc/$h$
and the field of view of $5$ deg on a side.

Within the \textsc{MapSim} code we have recently implemented also the
possibility to construct a corresponding light-cone of haloes and
subhaloes that resemble the underlying randomisation of the associated
matter density distribution along the
line-of-sight. Friends-of-Friends groups, $M_{\rm 200}$-haloes and
subhaloes are subdivided according to the various constructed planes;
for each of them we compute the corresponding redshift from their
comoving distance from the observer and their angular position in the
sky with respect to the assumed field of view. In order to avoid edge
effects when re-constructing the lensing properties from virialized
structures, we extracted haloes and subhaloes from a field of view
$2.5$ deg larger on each side. This means that haloes and subhaloes
are extracted from a region of $10 \times 10$ sq. degrees, centered in
the same sky position as the cone from which we extract the
particles. Halo and subhalo catalogues are saved in complementary
files with respect to the corresponding lens planes. We highlight that
in order not to double-count the mass in haloes we do not consider the
main subhalo within the \textsc{subfind} catalogues which typically
account for the smooth halo component.  As an example in the right
panel of Fig.~\ref{figMapandFoF} we plot on top the convergence map,
the positions of the FoF groups more massive than $5\times
10^{13}\mathrm{M_{\odot}}/h$ within the light-cone from $z=0$ to
$z_s=4$. The various size coloured circles refer to different masses
as indicated in the label.

\begin{figure*}
\includegraphics[width=0.275\hsize]{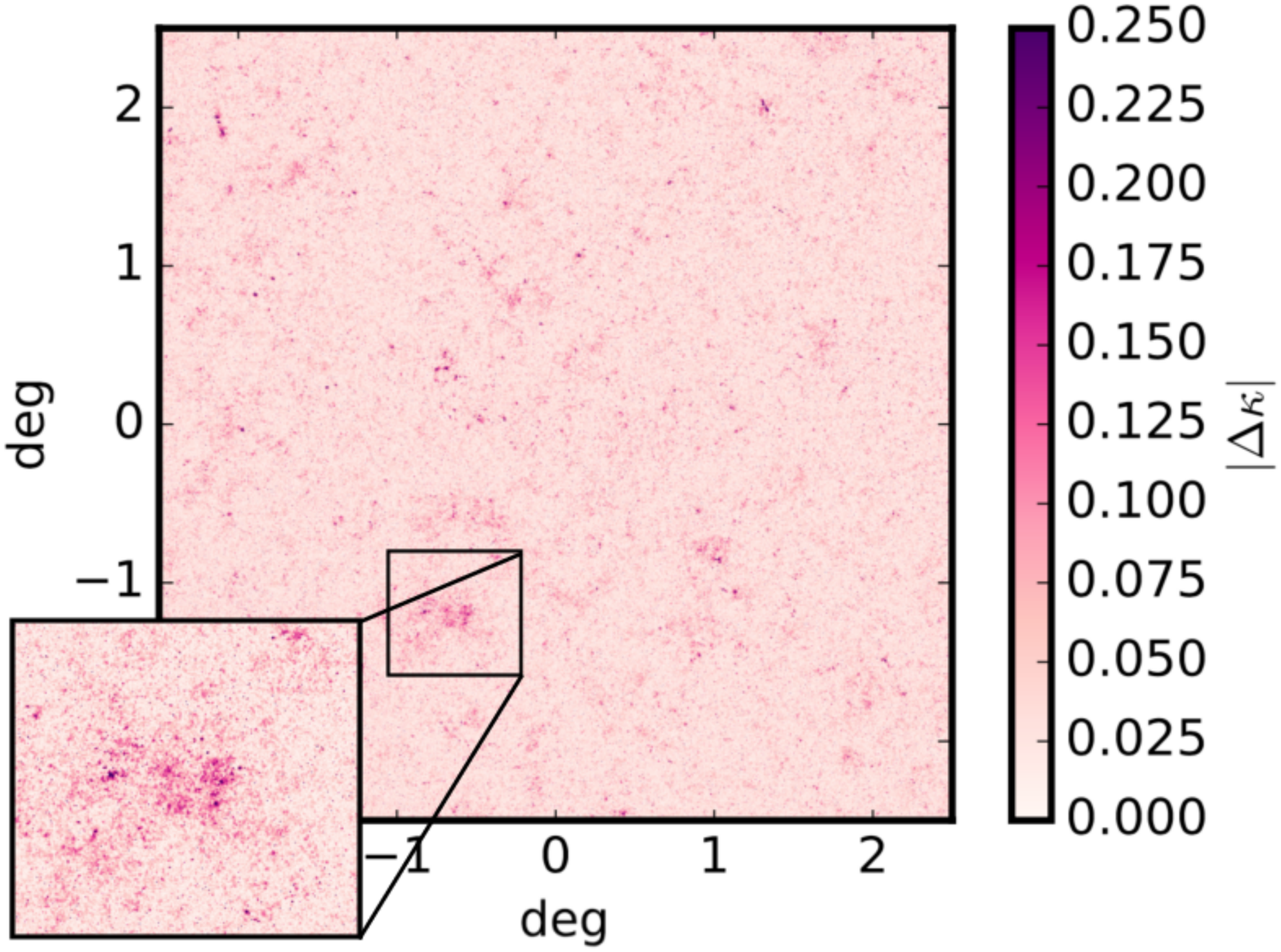}
\includegraphics[width=0.31\hsize]{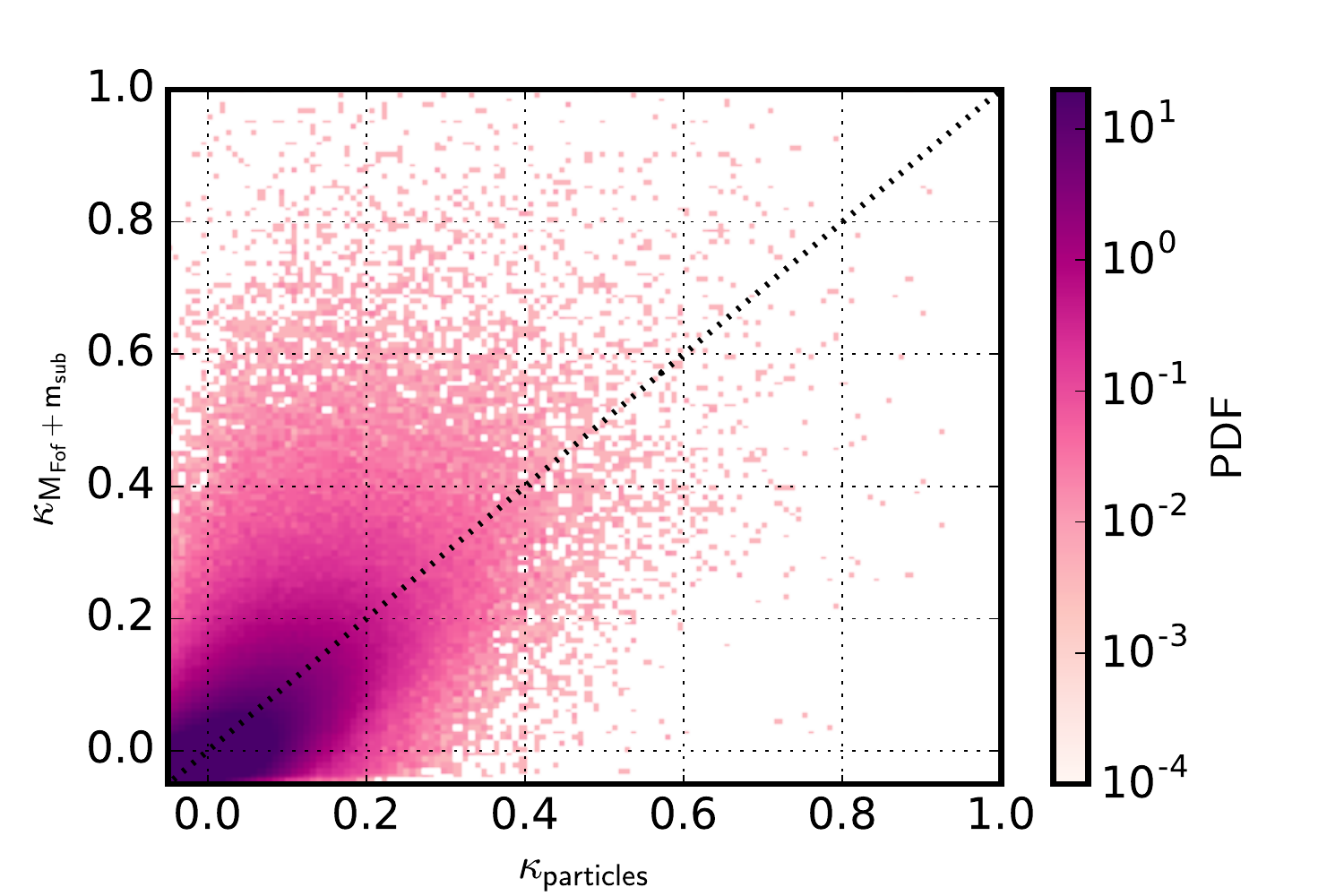}
\includegraphics[width=0.31\hsize]{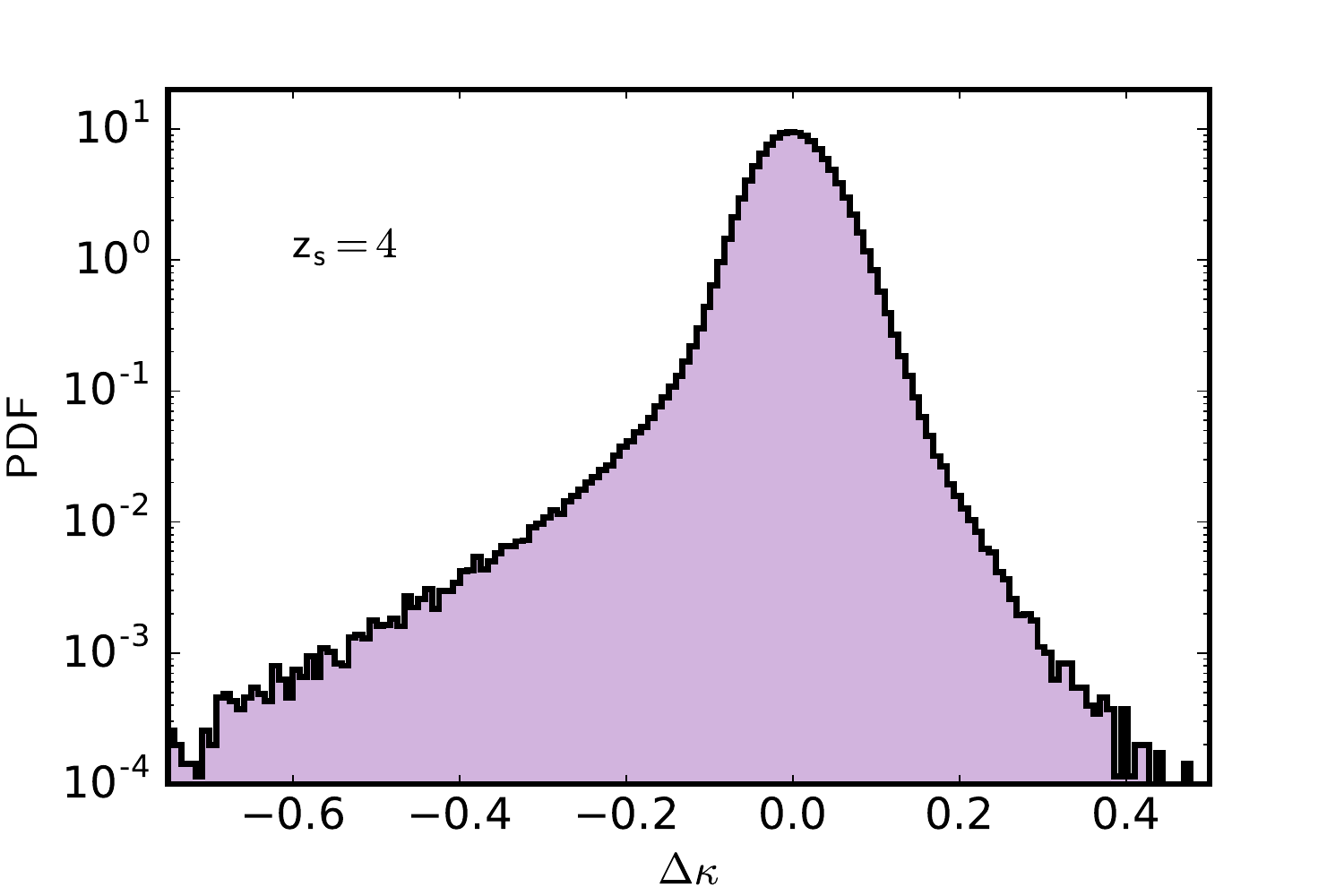}
\caption{Left panel: absolute difference between the convergence map
  computed using particles and the one using FoF-haloes plus
  subhaloes. Central panel: pixel by pixel correlation between the two
  maps.  Right panel: probability distribution function of the
  difference between the two maps.\label{figDiff}}
\end{figure*}

\section{A Weak Lensing Halo Model approach: \textsc{WL-MOKA}}
\label{sechmmodel}
The different statistical analyses performed in the last twenty-years
on the post-processing data of various numerical simulations have
given the possibility to reconstruct in good details the dark matter
halo structural properties over a wide range of masses
\citep{springel01b,gao04,giocoli08b}. In particular, many works seem
to converge toward the idea that virialized haloes tend to possess a
well defined density profile
\citep{navarro96,moore98,rasia04}. Following the \citet{navarro96}
(hereafter NFW) prescription we assume the density profile of haloes
to follow the relation:
\begin{equation}
  \rho(r|M_{\rm h}) = \frac{\rho_s}{(r/r_s)(1+r/r_s)^2}\,,  
\end{equation}
where $r_s$ is the scale radius, defining the concentration $c_{\rm h}
\equiv R_{\rm h}/r_s$ and $\rho_s$ the dark matter density at the
scale radius:
\begin{equation}
  \rho_s = \frac{M_{\rm h}}{4 \pi r_s^3}
  \left[ \ln(1+c_{\rm h}) - \frac{c_{\rm h}}{1+c_{\rm h}}\right]^{-1}\,,
\label{eqrhos}
\end{equation}
$R_{\rm h}$ is the radius of the halo which may varies depending on
the halo over-density definition. In this analysis we will adopt ($i$)
the mass inside the Virial radius for the FoF groups:
\begin{equation}
  M_{\rm vir}         =          \frac{4         \pi}{3}         R_{\rm vir}^3
  \frac{\Delta_{\rm vir}}{\Omega_{m}(z)} \Omega_0 \rho_c\,,
\label{massdef}
\end{equation}
and ($ii$) the mass inside a sphere enclosing $200$ times the critical matter density
$\rho_c(z)$ of the Universe:
\begin{equation}
  M_{\rm 200} = \frac{4 \pi}{3} R_{\rm 200}^3 200 \frac{\Omega_0}{\Omega_{\rm m}(z)}
  \rho_c\,,
\label{massdef200}
\end{equation}
where $\Omega_0 \equiv \Omega_{\rm m}(0)$ represents the matter
density parameter at the present time and $\Delta_{\rm vir}$ is the
virial over-density \citep{eke96,bryan98}, $R_{\rm vir}$ and $R_{\rm
  200}$ symbolise the virial and the $200$ critical radius of the
halo, that is the distance from the halo centre that encloses the
desired density contrast; $\rho_c$ represents the critical density at
the present time.

The halo concentration $c_{\rm h}$ is a decreasing function of the
host halo mass. This relation is explained in terms of hierarchical
clustering within CDM-universes and of different halo-formation
histories \citep{vandenbosch02,deboni16}.  Small haloes form first in
a denser universe and then merge together forming the more massive
ones: galaxy clusters sit at the peak of the hierarchical pyramid
being the more recent structures to form
\citep{bond91,lacey93,sheth04a,giocoli07}. This trend is reflected in
the mass-concentration relation: at a given redshift smaller haloes
are more concentrated than larger ones. Different fitting functions
for numerical mass-concentration relations have been presented by
various authors \citep{bullock01a,neto07,duffy08,gao08}. In this work,
we adopt the relation proposed by \citet{zhao09} which links the
concentration of a given halo with the time $t_{0.04}$ at which its
main progenitor assembles $4$ percent of its mass. \citet{giocoli12b}
have found that this relation works very well for virialized masses
$M_{\rm vir}$ while the parameters of the model need to be slightly
modified for the $M_{\rm 200}$ definition \citep{giocoli13}.  We want
to underline that the model by \citet{zhao09} also fits numerical
simulations with different cosmologies; it seems to be of reasonably
general validity within few percents of accuracy. For the mass
accretion history model of the two mass over-density definitions
($M_{\rm vir}$ or $M_{\rm 200}$) we adopt the relations by
\citet{giocoli12b} and \citet{giocoli13}. Those models are quite
universals and give the possibility to generalise the relations
eventually also to non-standard models \citep{giocoli13}. In
particular, the concentration mass relation mainly impacts on the
behaviour of the power spectrum at scales below $1\,h^{-1}$Mpc
as discussed in details by \citet{giocoli10b}.

Due to different assembly histories, haloes with same mass at the same
redshift may have different concentrations
\citep{navarro96,jing00,wechsler02,zhao03a,zhao03b}.  At fixed host
halo mass, the distribution in concentration is well described by a
lognormal distribution function with a rms $\sigma_{\ln c}$ between
$0.1$ and $0.25$ \citep{jing00,dolag04,sheth04b,neto07}. In this work
we adopt a lognormal distribution with $\sigma_{\ln c}=0.25$.

In our numerical simulation, subhaloes have been identified using the
\textsc{subfind} algorithm. For the mass density distribution in
subhaloes we adopt the truncated Singular Isothermal Sphere (hereafter
tSIS) profile. This model accounts for the fact that the subhalo density
profiles are modified by tidal stripping due to close interactions
with the main halo smooth component and to close encounters with other
clumps, gravitational heating, and dynamical friction.  Such events
can cause the subhaloes to lose mass, and may eventually result in
their complete disruption \citep{hayashi03,choi07}.  We model the dark
matter density profile in subhaloes as \citep{keeton03},
\begin{equation}
  \rho_{\rm sub}(r) = \left\{ \begin{array}{ll}
    \dfrac{\sigma_v^2}{2 \pi G r^2} &r\le R_{sub}, \\
    0  &r>R_{sub} \\
  \end{array} \right.
\end{equation}
with  velocity dispersion  $\sigma_v$, and $R_{sub}$ defined as:
\begin{eqnarray}
m_{sub} &=& \int_0^{R_{sub}} 4 \pi r^2 \rho_{sub}(r) \mathrm{d}r \Rightarrow \nonumber \\
R_{sub} &=& \dfrac{G\,m_{sub}}{2\,\sigma_v^2}\,. \label{rsub}
\end{eqnarray}
To compute the velocity dispersion we use the same implementation
described in the \textsc{MOKA} code by \citet{giocoli12a}. The tSIS
profile well represents galaxy density profiles on scales relevant for
strong lensing. Previously, different authors have used this model to
characterise the lensing signal by substructures after stripping
\citep{metcalf01}. \\ In Table~\ref{tabsum} we summarise the halo
model properties which we use to construct the convergence maps using
our algorithm.

\begin{table*}
  \caption{\label{tabCases}Summary of the halo and subhalo properties
    considered in our models when building the effective convergence
    maps\label{tabsum}}
  \begin{tabular}{ l |  c | c }
    Case & c-M relation & profile \\ \hline \hline
    FoF & \citet{zhao09} & NFW \\ 
    M200 & \citet{giocoli13} & NFW \\ 
    FoF+Subs & \citet{zhao09} & NFW (haloes)+ tSIS (subhaloes) \\ \hline \hline

\end{tabular}
\end{table*}

Assuming spherical symmetry for the matter density profile in haloes,
we can compute the surface mass density $\Sigma(x_1,x_2)$ associated
with a density profile $\rho(r)$ extending up to the virial radius
$R_{\rm vir}$ as:
\begin{equation}
\Sigma(x_1,x_2) = 2 \int_0^{R_{\rm vir}} \rho(x_1,x_2,\zeta) \mathrm{d} \zeta,\label{eqsigma}
\end{equation}
where $x_1$, $x_2$ and $\zeta$ represents the three-dimensional
coordinates and $r^2 = x_1^2 + y_2^2 + \zeta^2$; this quantity is then
used to define the convergence as in eq.~(\ref{eqconvergence}).

As described by \citet{bartelmann96a} the Navarro-Frank-White density
profile has a well defined primitive for the integral in
equation~(\ref{eqsigma}) and its convergence can be derived
analytically, as well as for the tSIS profile.

\begin{figure*}
  \includegraphics[width=0.3\hsize]{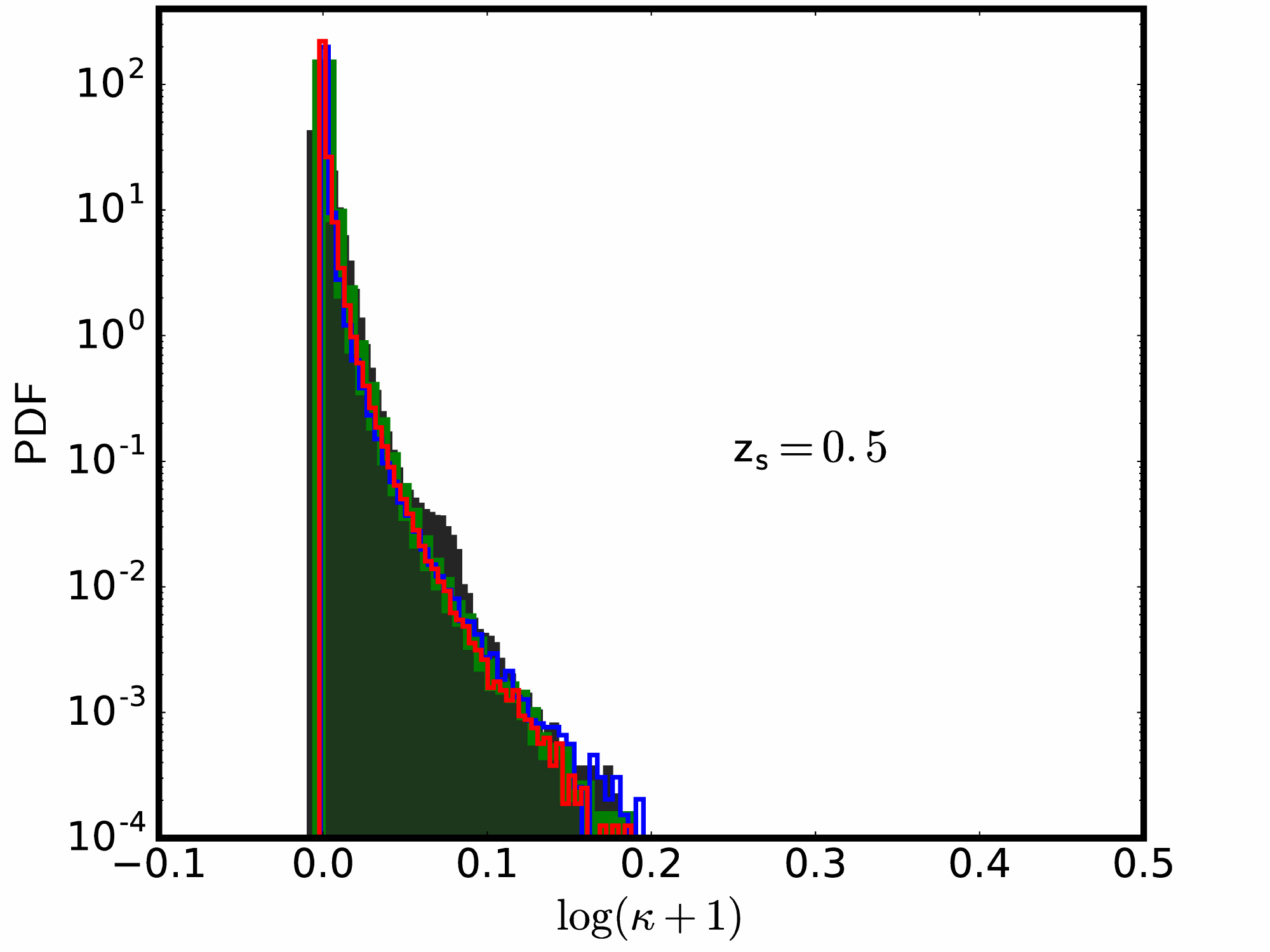}
  \includegraphics[width=0.3\hsize]{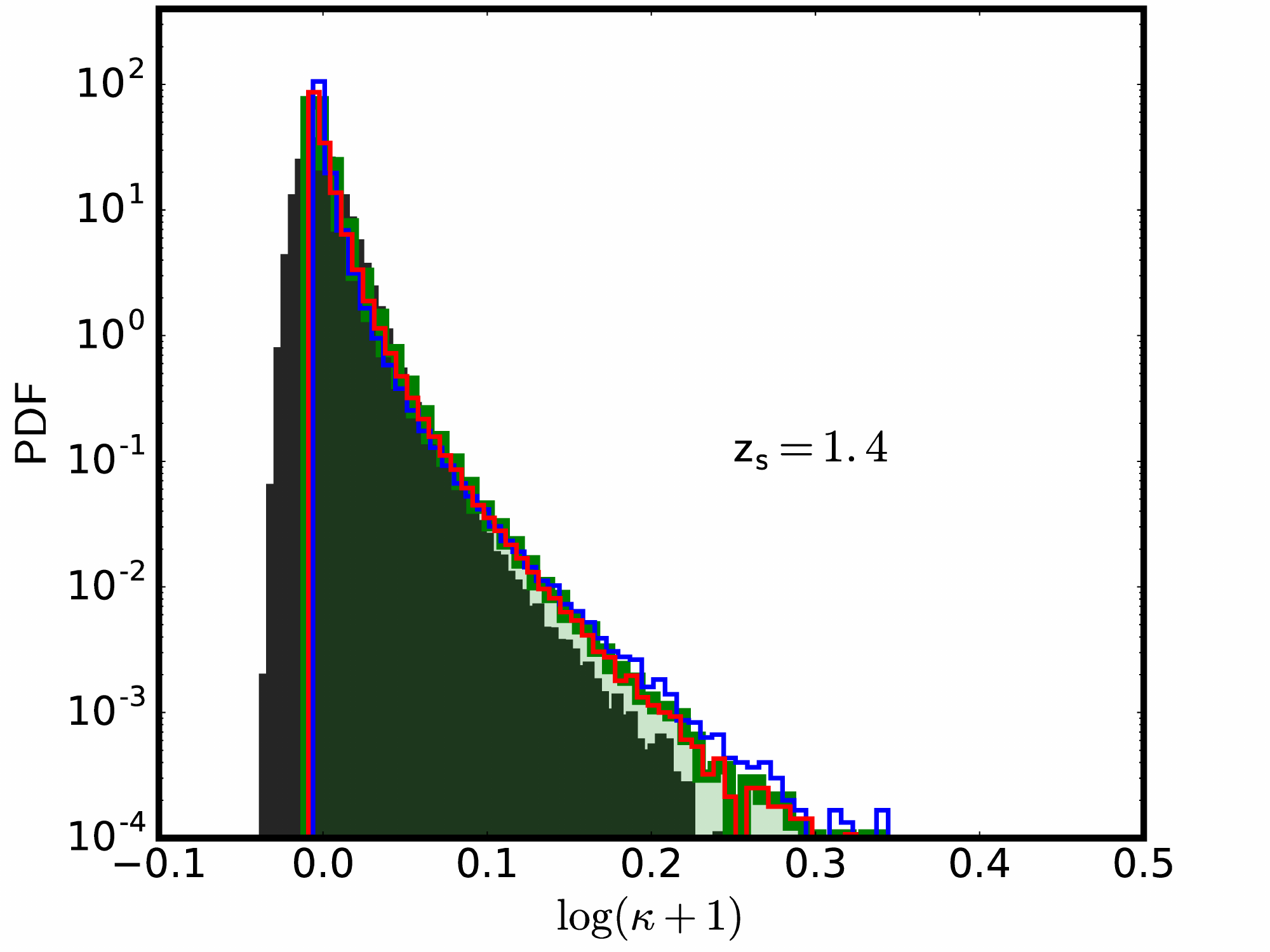}
  \includegraphics[width=0.3\hsize]{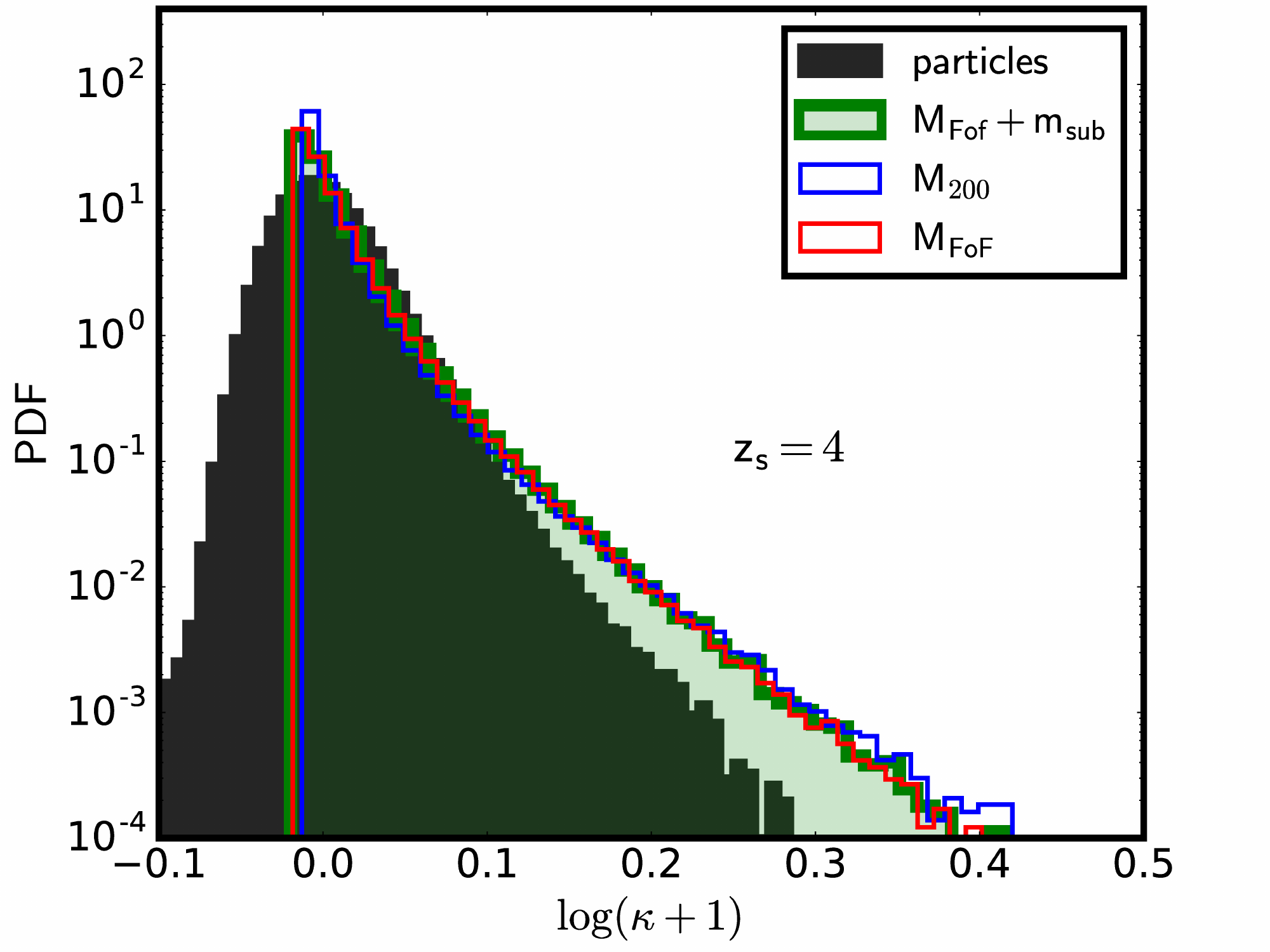}
  \caption{Probability Distribution Functions (PDF) of the convergence
    maps presented in Fig.~\ref{figmaps}.  Left, central and right
    panels show the PDFs for three different source redshifts:
    $z_s=0.5$, $1.4$ and $4$, respectively. The black histograms show
    the PDF for the convergence map constructed using the
    \textsc{glamer} ray-tracing pipeline, while the blue and red are
    the ones computed from the $M_{\rm 200}$ and $M_{\rm FoF}$ halo
    catalogues. The green histograms show the distribution function of
    the convergence map where the Friends-of-Friends haloes contain
    also substructures.  \label{pdf00fig}}
\end{figure*}  

In Fig.~\ref{figmaps}, we show 4 convergence maps of the same
light-cone realisation extending up to redshift $z_s=4$.  In the top
left panel (in black scale), we created the convergence map using
ray-tracing in the light-cone constructed from the particles extracted
from the simulation snapshots. In the top right (in red scale) and
bottom left (in blue scale) we present the convergence maps
constructed using the $M_{\rm FoF}$ and the $M_{\rm 200}$ halo
catalogues, respectively. By eye it is possible to spot that using the
halo catalogues the overall surface mass density distribution is quite
well traced. However it is noticeable with more careful analysis that the map constructed
using the Friends-of-Friends catalogue presents much more clustering
of low mass haloes. This is be due ($i$) to numerical resolution of
the simulation: FoF haloes with less than 10 particles within $200$
times the critical density are not well resolved and not stored in the
corresponding $M_{\rm 200}$ catalogue and ($ii$) to the possible non
universality of the mass function defined with $M_{\rm 200}$ haloes
\citep{tinker08,despali16}.  In general it is interesting to notice
that $M_{\rm FoF}$ and $M_{\rm 200}$ contain typically a different
fraction of the total mass in the simulation. Using the relations
calibrated from numerical simulations by \citet{despali16}, we notice
assuming the same mass resolution -- down to ten dark matter particles
-- and box-size of our reference run, at $z=0$. The mass contained in
$M_{\rm FoF}$ haloes is approximately $30\%$ of the total mass in the
simulation, while in $M_{\rm 200}$ haloes it is less than $25\%$; at
$z=1$ the two fractions become $15\%$ and $12\%$, respectively, while
at $z=3$ they are both approximately $1.5\%$, since the two mass
over-density definitions get closer and closer at high redshifts
\citep{eke96,bryan98}.

Convergence maps constructed by summing the surface mass density
contribution of all haloes present within the halo catalogues, and
weighting them with the critical surface density as in
eq.~(\ref{eqconvergence}), are effective convergence maps and are
forced to have an average value of the convergence $\bar{\kappa} =
\dfrac{\sum_{i=1}^{N_{pix}}\kappa_i}{N_{pix}}=0$.  This implies that
conservatively each convergence map describes the perturbed matter
density distribution with respect to an average background value. We
underline also that this point is important when we construct the
effective convergence maps using only haloes or using both haloes and
subhaloes; in order not to over-count the masses in both cases the
average value of the convergence in each constructed plane is set to
be zero. This kind of approach has also been used in constructing the
convergence map implying the full ray-tracing technique -- as in the
top left panel of the figure: since the rays are propagated between
planes using the standard distances in a Robertson-Walker metric which
assumes a uniform distribution of matter the addition of matter on
each of the planes will, in a sense, over-count the mass in the
universe.  Without correcting for this, the average convergence from
the planes will be positive and will cause the average distance for a
fixed redshift to be smaller than it should be. To compensate for the
contained density between the planes, the ensemble average density on
each plane is subtracted. Each plane then has zero convergence on
average and the average redshift-distance relation is as it would be
in a perfectly homogeneous universe.  Finally, the bottom right panel
of Fig.~\ref{figmaps} (in green scale) presents the convergence map
constructed using the FoF haloes plus the subhaloes. In this case
comparing this map with respect to the one in red scale, where we use
only the FoF haloes, we notice an increase of small scale
perturbations.  In Fig.~\ref{figDiff} we display the statistical
difference between the maps computed using particles and FoF-haloes
plus subhaloes. In the left panel we show the absolute difference map
between the two cases. The central panel exhibits the pixel by pixel
correlation between the two maps, while the left panel presents the
Probability Distribution Function (PDF) of the difference $\Delta
\kappa = \kappa_{\rm particles} - \kappa_{\rm M_{FoF}+m_{sub}}$. From
the figures what is mainly appearing is that the effective convergence
map computed using haloes and subhaloes mainly trace the matter
density distribution on small scales where non-linear structures and
clumps are present, however still differences appear mainly due to
projection effects, filamentary structures -- as better displayed in
the small panel in the left figure -- and sheets.

\subsection{Probability Distribution Function of the Convergence Fields}

To quantify the previous discussion, in Fig.~\ref{pdf00fig} we display
the Probability Distribution Function of the convergence maps
presented in Fig.~\ref{figmaps}. Left, central and right panels show
the PDF of the convergence constructed for sources at $z_s=0.5$, $1.4$
and $4$, respectively. Black, red, blue and green coloured histograms
show the four corresponding cases used to construct the convergence
map: particles, FoF groups, $M_{\rm 200}$-haloes, and FoF groups with
subhaloes. From the panels in the figure, we notice that for $z_s=0.5$
the four histograms are very similar and that the inclusion of
substructures creates some pixels with larger convergence values which
may correspond to the core of clumps.  In the central and right panels
we notice that the PDF of the convergence map constructed using the
particles does not present pixels with convergence $\kappa \gtrsim
0.75$, this is probably due to the numerical and force resolution of
the simulation which does not permit to resolve with a reasonable
number of particles the cores of haloes and subhaloes. In addition, the
black histograms display distinct tails with negative
convergence. This is probably due to the sampling of the matter
density distribution that is not bound to haloes -- and that we are
missing in our halo modelling formalism. We will discuss more about
that in the next sections.

\subsection{Building up the convergence power spectra}

Following the halo model formalism, the matter density distribution in
the universe is assumed to be associated to virialized haloes
\citep{cooray02}. The mean density within the Universe can so be
computed from the relation:
\begin{equation}
\bar{\rho} = \int m \; n(m) \; \mathrm{d}m,
\end{equation}
where $n(m)$ represents the halo mass function. The three-dimensional
matter power spectrum can be then decomposed in:
\begin{equation}
P_{\rm \delta}(k,z) = P_{\rm 1h}(k,z) + P_{\rm 2h}(k,z),\label{eqhalomodel}
\end{equation}
where $P_{\rm 1h}(k)$ represents the power spectrum of the matter
density distribution within one halo, while $P_{\rm 2h}(k)$ describes
the power spectrum of the matter density distribution between two
distant haloes. The two terms can be read as:
\begin{eqnarray}
P_{\rm 1h}(k,z) =  &&\int \left( \dfrac{m}{\bar{\rho}}\right)^2 n(m,z) u^2(k|m) \mathrm{d}m  \label{eq1halo}\\
P_{\rm 2h}(k,z) = &&\int \left( \dfrac{m_1}{\bar{\rho}}\right) n(m_1,z) u(k|m_1) \mathrm{d}m_1 \\ 
	&&\int \left( \dfrac{m_2}{\bar{\rho}}\right) n(m_2,z) u(k|m_2) \mathrm{d}m_2 P_{\rm h h}(k|m_1,m_2),\nonumber
\end{eqnarray}
where $u(k|m)$ represents the Fourier transform of the dark matter
density profile and $P_{\rm h h}(k|m_1,m_2)$ describes the halo-halo
power spectrum that can be expressed in terms of the halo-matter bias
parameter $b(m)$ and the linear matter power spectrum $P_{\delta,\rm
  lin}(k)$:
\begin{equation}
P_{\rm h h}(k|m_1,m_2) = b(m_1) b(m_2) P_{\delta,\rm lin}(k).
\end{equation}
Including the presence of substructures within haloes adds more
equations to the halo model that can be trivially solved considering
the correlation between the smooth and the clump components both
within the 1-halo and the 2-halo term \citep{sheth03c,giocoli10b}.

\begin{figure*}
  \includegraphics[width=0.33\hsize]{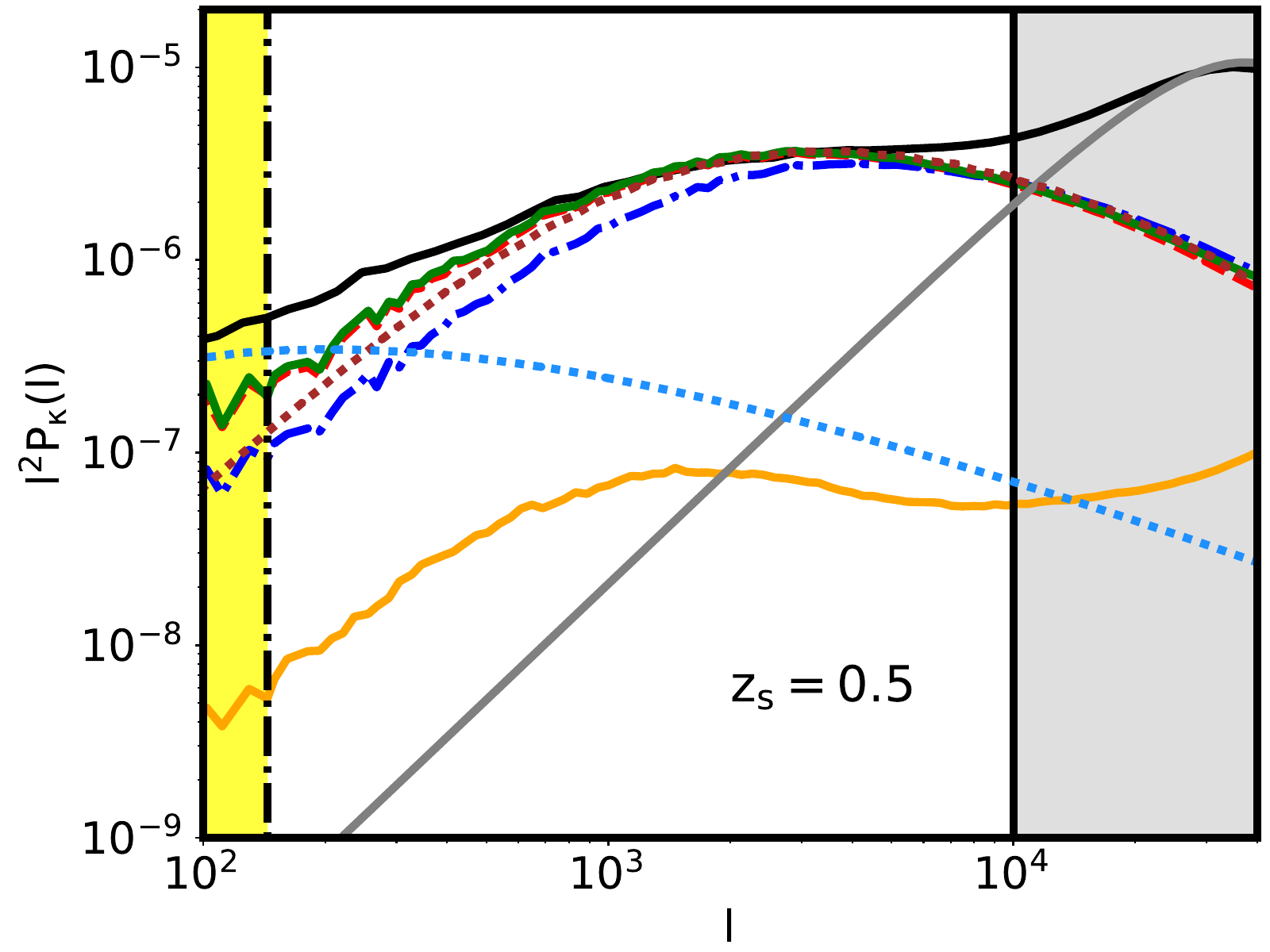}
  \includegraphics[width=0.33\hsize]{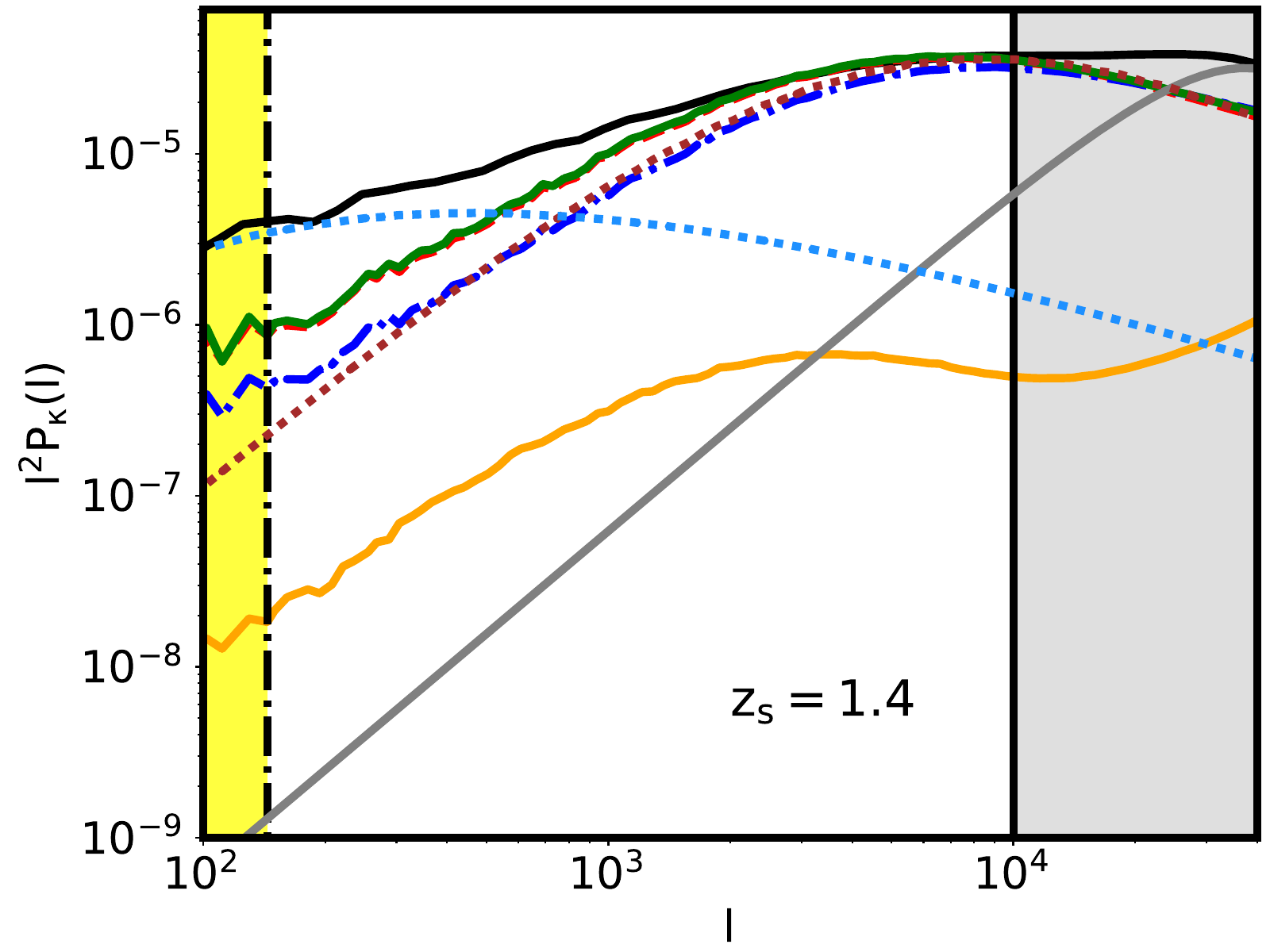}
  \includegraphics[width=0.33\hsize]{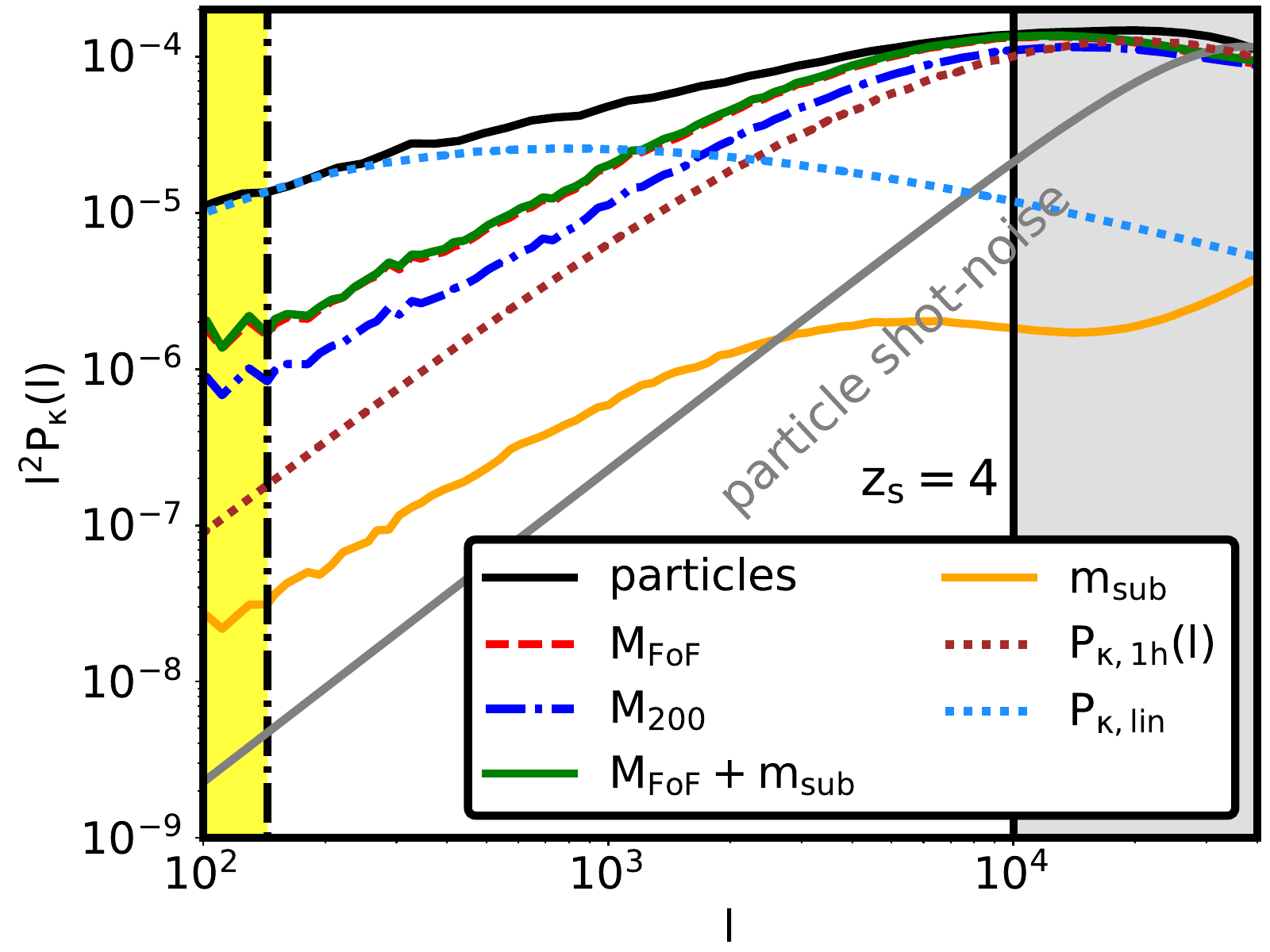}    
  \caption{Convergence power spectra averaged over $25$ different
    realisations of the light-cones considering sources at three fixed
    source redshifts, from left to right we have $z_s=0.5$, $z_s=1.4$
    and $z_s=4$. The black curves show the power spectra computed from
    the ray-tracing pipeline including all the particles in the
    light-cone extracted from the numerical simulation (as described
    in \citet{giocoli15}). The red dashed and blue dot-dashed curves
    show the results of the two considered halo catalogues $M_{\rm
      FoF}$ and $M_{\rm 200}$, while the solid orange lines present
    the contribution of the substructures.  The solid green curves
    display the total convergence power spectra of Friends-of-Friends
    haloes and subhaloes. The grey curves describe the shot-noise
    contribution to the power spectra computed from the ray-tracing
    simulations using particles. \label{figpowerspectra1}}
\end{figure*}  

The convergence power spectrum, to first order, can be expressed as an
integral of the three-dimensional matter power spectrum computed from
the observer looking at the past lightcone from the present epoch up
to a given source redshift \citep{bartelmann01}. In this approximation
it is assumed that the light rays travel along unperturbed paths and
all terms higher than first order in convergence and shear can be
ignored.  Defining $f(w)$ as the angular radial function, that depends
on the comoving radial coordinate $w$ given the curvature of the
universe, we can write the convergence power spectrum at a given
source redshift $z_s$ -- with a corresponding radial coordinate $w_s$
-- as:
\begin{equation}
P_{\kappa}(l) = \dfrac{9 H_0^4 \Omega_m^2}{4 c^4} \int_0^{w_s(z_s)}
\dfrac{f^2(w_s-w)}{f^2(w_s)a^2(w)} P_{\rm \delta}\left(\dfrac{l}{f(w)},w\right) \, \mathrm{d}w.
\label{eqborn}
\end{equation}
Analogously from the constructed effective convergence maps we can
compute the corresponding power spectrum as:
\begin{equation}
 \langle \hat{\kappa}(\mathbf{l}) \hat{k^*}(\mathbf{l}') \rangle = {4 \pi^2}
 \delta_D(\mathbf{l}-\mathbf{l}') P_{\kappa}(l),\label{pkmapeq}
\end{equation}
where $\delta_D^{(2)}$ represents the Delta Dirac in two dimensions.

In Fig.~\ref{figpowerspectra1} we present the average power spectrum
of $25$ different light-cone realisations for three different source
redshifts: $z_s=0.5$, $z_s=1.4$ and $z_s=4$, from left to right
respectively. In each panel, the black curves display the spectra
computed using the ray-tracing pipeline on the particle distribution
and the grey curves show the associated particle shot-noise
\citep{vale03,giocoli16a}.  The shaded grey area marks the region
where the shot-noise term of the particles starts to dominate the
cosmic shear measurements, while the yellow shaded region indicates
the part below the angular Nyqvist mode sampled by our field of
view. Red dashed and blue dot-dashed curves show the power spectra
computed using the FoF and the $M_{\rm 200}$ haloes present within the
light-cones. The orange curves describe the contribution of the
subhaloes while the green solid curves exhibit the total contribution
of the Friends-of-Friends haloes and their associated subhaloes. From
the figure we can observe that the large scale behaviour of our
halo model power spectra manifests less power than expected from
linear theory (dotted light-blue curves).  The magenta dashed curves
display the one-halo term contribution of the analytical halo model as
in eq.~(\ref{eq1halo}) where we have integrated the theoretical mass
function \citep{sheth99b} from the minimum halo mass that we have in
the simulation $M_{\rm min}\approx 2.07\times 10^{12}M_{\odot}/h$ for
consistency. From the figure we notice that our halo model for weak
lensing captures quite well the 1-halo term plus the one related to
the matter between haloes, but misses the linear contribution of
matter distributed among haloes; that is matter density fluctuations
that are not attached to non-linear structures, and possibly tracing
sheets and filaments.

\begin{figure}
  \includegraphics[width=\hsize]{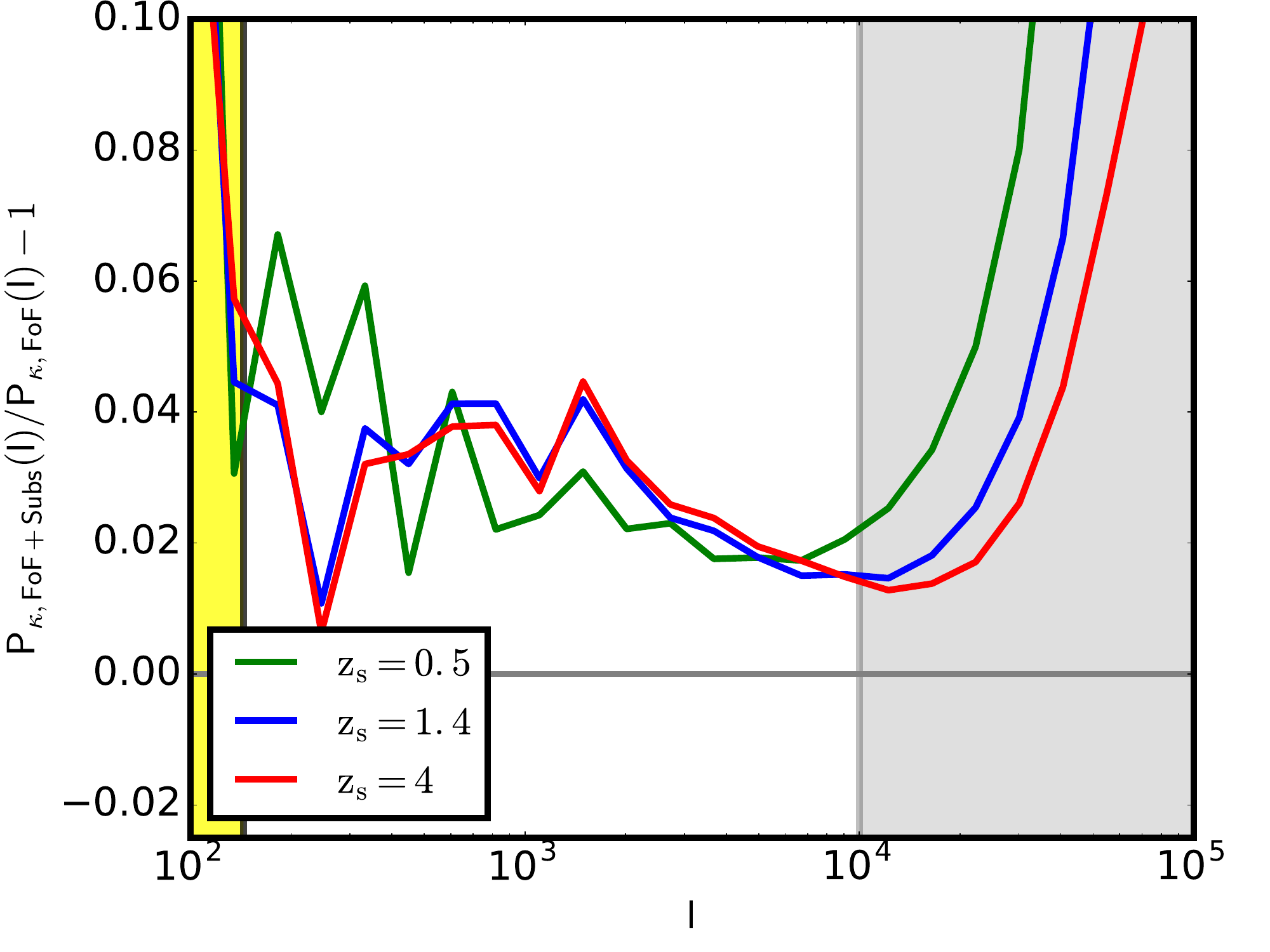}
  \caption{Relative contribution to the cosmic shear power spectrum of
    the subhaloes with respect to the FoF-haloes for three different
    source redshifts: $z_s=0.5$ (green), $z_s=1.4$ (blue) and $z_s=4$
    (red).  \label{figsubhaloes}}
\end{figure}

The relative contribution of subhaloes to the power spectrum with
respect to the smooth component is displayed in
Fig.~\ref{figsubhaloes}. The green, blue and red curves represent the
subhalo contribution for three different source redshifts. From the
figure, we notice that typically subhaloes contribute to approximately
$3\%$ to the convergence power spectrum and that their contribution
becomes significant for scales below $5$ arcmin, which correspond to
approximately is $l \approx 10^4$. In particular those scales are not
well resolved within the numerical simulation due to particle and
force limitations while well described by our halo model
formalism. 
We remind the reader that in those regimes, a consistent
treatment of the baryonic contribution is very critical
\citep{harnois-deraps15}, and this will be addressed in an upcoming
paper (Giocoli, Monaco et al. in preparation).

\subsection{Effective linear contribution to the weak lensing halo model}
\begin{figure}
  \includegraphics[width=\hsize]{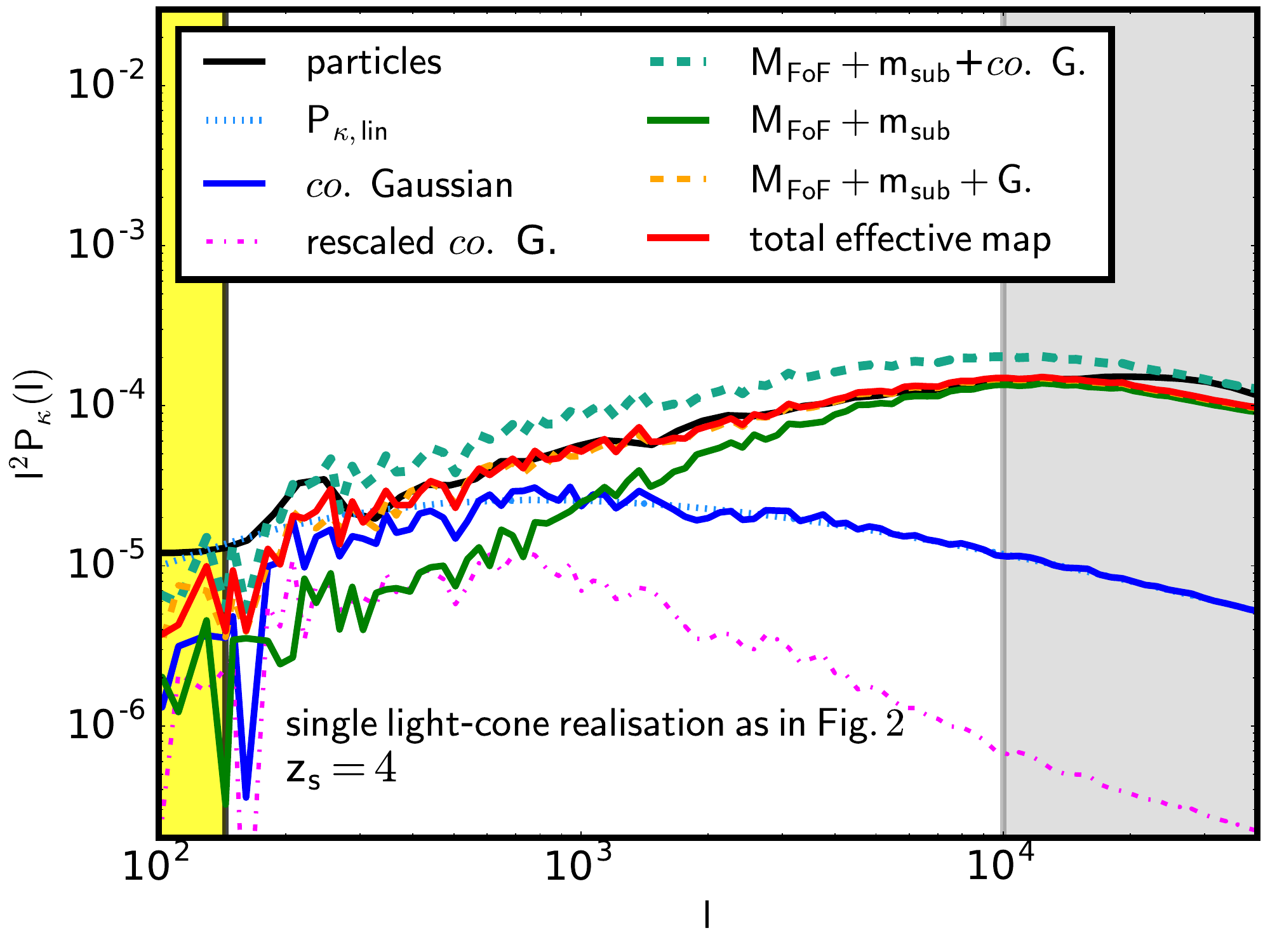}
  \caption{Convergence power spectrum for sources at $z_s=4$. The
    dotted cyan curve shows the predictions using the linear matter
    power spectrum while the blue line shows the power spectrum of a
    random Gaussian realisation. The green curve displays the
    contribution to the power spectrum arising from the haloes
    $P_{\kappa_{\rm hm}}(l)$ and the dashed orange displays
    $P_{\kappa_{\rm hm}}(l) + P_{\kappa_{\rm lin}}(l)$, the sum of the
    halo contribution and the linear Gaussian realisation
    power-spectra. The black solid line shows the power spectrum of
    the ray-tracing pipeline using particles from the simulation,
    while the light-blue dotted line displays the power spectrum of
    the convergence map calculated summing $\kappa_{\rm hm}$ and
    $\kappa_{\rm lin}$, a map that is coherent ($co.$) in phase with
    the halo population. The solid red curve shows the effective total
    map where the amplitude of $\kappa_{\rm lin}$ is rescaled
    according to eq.~(\ref{eqampl}) -- see the text for more
    details.\label{fignonlinearkappa}}
\end{figure}

As discussed in the previous section, the halo model formalism we have
implemented so far is missing the effective contribution of the linear
matter density distribution presents among the haloes, which may be
tracing sheets and filaments. Recently
  \citet{vandaalen15}, using a set of cosmological numerical
  simulations, discussed how much non-virialized matter contributes to
  the total matter power spectrum.  In particular they showed that the
  larger the region around the virialized haloes that is included, the
  larger the halo contribution to the matter power spectrum will be.
  The matter power spectrum of haloes for $3 < k <
  100\,h\mathrm{Mpc^{-1}}$ enclosing $200$ times the critical density
  is smaller then that enclosing $200$ times the background and, in
  turn, of that of the mass residing within the FoF groups.  Going
  from three to two dimensions it can be noticed from the panels
present in Fig.~\ref{figpowerspectra1},  our model
properly reconstructs the 1-halo term plus a 2-halo-like term but has
less power at large scales with respect to the ray-tracing power
spectrum as computed using particles. Consistent with the
  results of \citet{vandaalen15}, we notice that the convergence power
  spectra of the matter in $M_{\rm 200}$ is smaller than that of the
  matter within the FoF groups. However the relative difference
  between the two depends on the considered source redshift: $M_{\rm
    200}$ has a pseudo-redshift evolution, as discussed by
  \citet{diemer13}, that depends on the evolution of the Hubble
  function with the cosmic time. To better clarify and understand the
  contribution of the matter in virialized haloes, in
Fig.~\ref{fignonlinearkappa} we display the convergence power spectrum
for sources at redshift $z_s=4$.  The black curve represents the power
spectrum from the ray-tracing simulation using particles for one
light-cone realisation, while the green curve displays our halo model
contribution from haloes and subhaloes. The cyan dotted line shows the
convergence power spectrum $P_{\kappa, \rm lin}(l)$ computed from the
linear theory  assuming $z_s=4$, while the blue curve
displays the power spectrum of a random Gaussian realisation
$\kappa_{\rm lin,r}$ of the theoretical linear cosmic shear power
spectrum $P_{\kappa, \rm lin}(l)$ in amplitude with a random phase
-- the subscript r stands for random in phase. The dashed
orange curve -- almost overlapping the red one -- presents the
convergence power spectrum of a map computed by summing our halo model
convergence map $\kappa_{\rm hm}$ -- halo and subhalo contribution --
with $\kappa_{\rm lin,r}$ calculated for $z_s=4$. Computing its power
spectrum, because the cross-terms are zero, we can read:
\begin{eqnarray}
  \langle \hat{\kappa}(\mathbf{l}) \hat{k^*}(\mathbf{l}') \rangle &=&
  \langle \reallywidehat{(\kappa_{\rm hm} + \kappa_{\rm lin,r})}(\mathbf{l})  \nonumber
  \reallywidehat{(\kappa_{\rm hm} + \kappa_{\rm lin,r})^*}(\mathbf{l}) \rangle \\ 
  &=&  4 \pi^2 \delta_D(\mathbf{l}-\mathbf{l}') \left(
  P_{\kappa_{\rm hm}}(l) + P_{\kappa_{\rm lin,r}}(l) \right),
\end{eqnarray}
where $\mathbf{l}\equiv(l_1,l_2)$, $P_{\kappa_{\rm hm}}(l)$ represents
the power spectrum using our halo model formalism and $P_{\kappa_{\rm
    lin,r}}(l)$ is the power spectrum of the Gaussian realisation of
the theoretical linear prediction with random phase. Finally, the
light-blue dashed curve shows the convergence power spectrum of a map
computed summing to $\kappa_{\rm hm}$ the map of a Gaussian
realisation of $P_{\kappa, \rm lin}(l)$ random in amplitude but with a
phase coherent (indicated with $co.$ in the figure) with the
structures present within $\kappa_{\rm hm}$. In order to construct a
map that is coherent in phase with the convergence map built from
haloes and subhaloes we define the Fourier transform of $\kappa_{\rm
  hm}$ as $\hat{\kappa}_{\rm hm}(l_1,l_2) \equiv
\mathrm{Re}\left[\tilde{\kappa}_{\rm hm}(l_1,l_2)\right]+ i\,
\mathrm{Im}\left[\tilde{\kappa}_{\rm hm}(l_1,l_2)\right]\equiv
\tilde{\kappa}_{\rm hm}(l_1,l_2) \cos[\phi(l_1,l2)] + i\,
\tilde{\kappa}_{\rm hm}(l_1,l_2) \sin[\phi(l_1,l_2)]$; we then
generate a Gaussian realization of the linear power spectrum with
amplitude $\tilde{\kappa_{\rm lin}}(l_1,l_2)$ and phase
\begin{equation}
\phi = \arctan \left( \frac{\mathrm{Im}\left[\tilde{\kappa}_{\rm
      hm}(l_1,l_2)\right]}{\mathrm{Re}\left[\tilde{\kappa}_{\rm
      hm}(l_1,l_2)\right]} \right).
  \end{equation}
This case is considered because we aim to ensure that the matter
present among virialized haloes is consistent with the non-linear
matter density distribution in a way to resemble sheets, filaments and
knots; moreover our aim is to develop a model which is independent of
the bias between halo and matter.  We stress also that we are aware
that adding together two fields that are coherent and computing the
power spectrum as in eq. (\ref{pkmapeq}) we obtain:
 \begin{eqnarray}
 \langle \hat{\kappa}(\mathbf{l}) \hat{k^*}(\mathbf{l}') \rangle &=& 
  \langle \reallywidehat{(\kappa_{\rm hm} + \kappa_{\rm lin})}(\mathbf{l}) 
  \reallywidehat{(\kappa_{\rm hm} + \kappa_{\rm lin})^*}(\mathbf{l}) \rangle \\ 
  &=&  4 \pi^2 \delta_D(\mathbf{l}-\mathbf{l}') \left(
  P_{\kappa_{\rm hm}}(l) + P_{\kappa_{\rm lin}}(l) + P_{\rm hm\otimes lin} 
  \right), \nonumber
 \end{eqnarray}
 where $P_{\kappa_{\rm hm}+\kappa_{\rm lin}}(l) = P_{\kappa_{\rm
     hm}}(l) + P_{\kappa_{\rm lin}}(l) + P_{\rm hm\otimes
   lin}$. $P_{\rm hm\otimes lin}$ indicates the cross-spectrum term
 between the two fields and that by definition $P_{\kappa_{\rm
     lin}}(l) = P_{\kappa_{\rm lin,r}}(l)$. From the figure we can
 notice that the normalisation of $P_{\kappa_{\rm hm}+\kappa_{\rm
     lin}}(l)$ is much higher than expected due to the cross-spectrum
 term between the two convergence maps that are in phase with each
 other where non-linear structures are present. In order to
 renormalize the computed power spectrum according to the expectation
 from linear theory, we define an effective linear power spectrum
 $P_{\kappa_{\rm eff,lin}}(l)$, with a phase coherent with the halo
 population, but with an amplitude renormalized according to the
 following relation:
 \begin{equation}
  A(l) = \dfrac{P_{\kappa,\rm lin}(l)}{P_{\kappa_{\rm hm}+\kappa_{\rm lin}}(l)}. \label{eqampl}
 \end{equation}
 The magenta dot-dashed curve in Fig.~\ref{fignonlinearkappa} shows
 the power spectrum of the effective linear map $\kappa_{\rm eff,lin}$
 that added to $\kappa_{\rm hm}$ gives our final result that is the
 total effective power spectrum displayed in red -- not far from the
 black curve as we will discuss in the next section.

\begin{figure*}
  \centering
  \includegraphics[width=0.45\hsize]{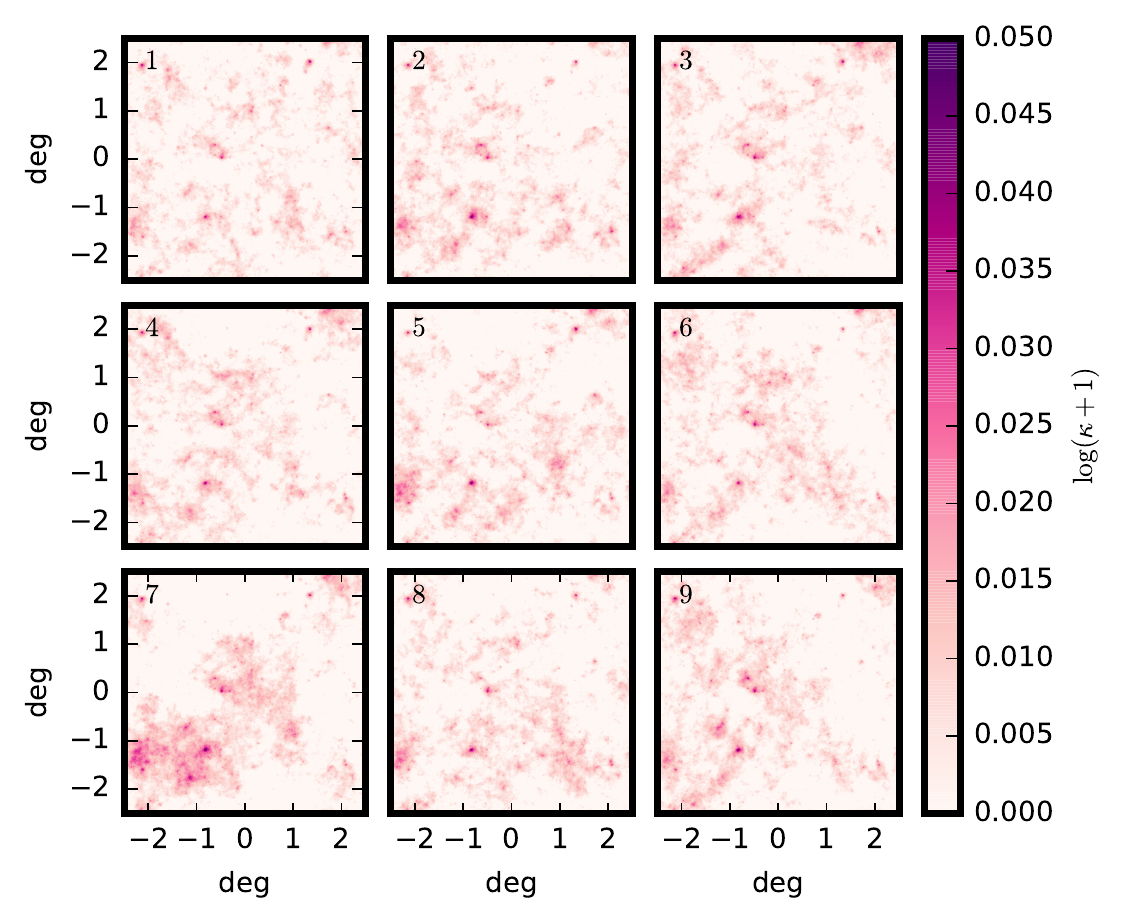}
  \includegraphics[width=0.45\hsize]{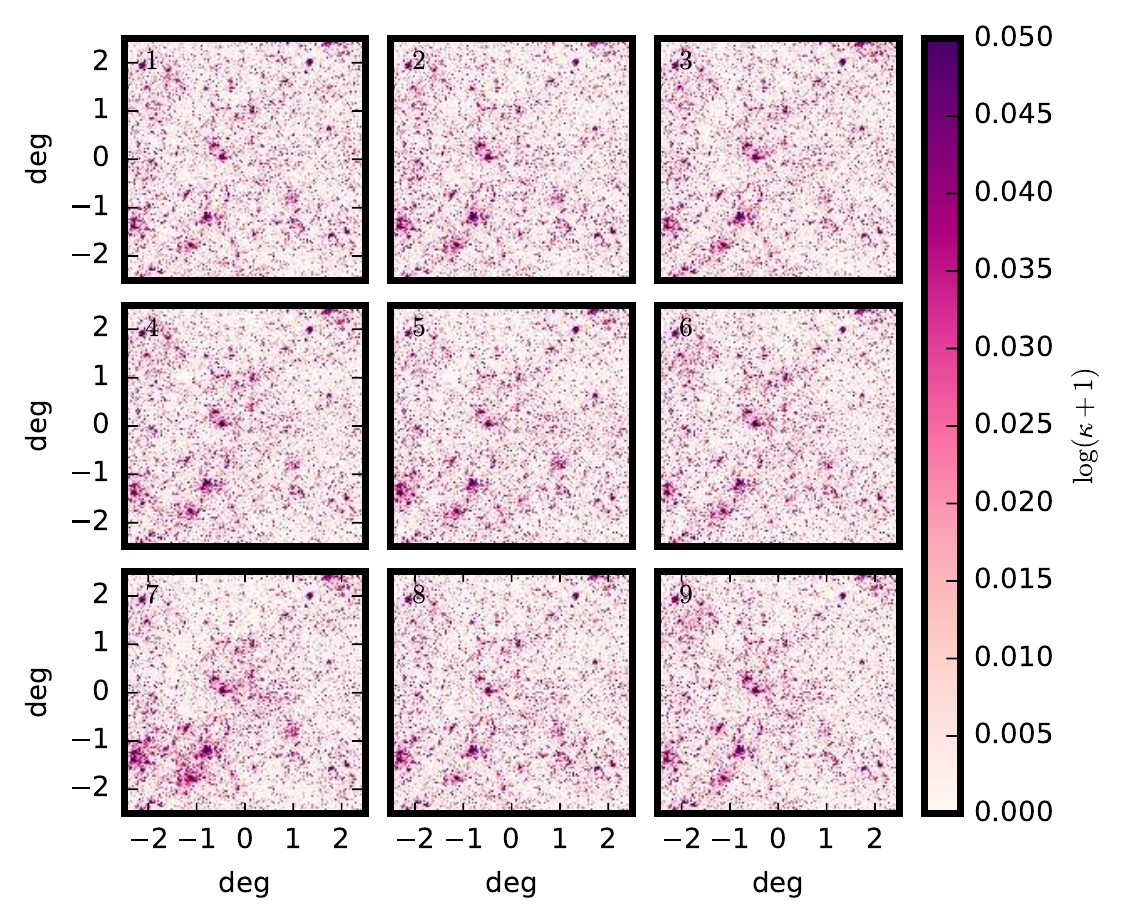}  
  \caption{Left panels: nine reconstructed effective linear
    convergence maps built from the theoretical linear predictions
    rescaled in amplitude as in eq.~(\ref{eqampl}). Their phase
    is consistent with the non-linear structures present in the
    field. Right panels: total effective convergence maps: $\kappa_{\rm
      hm} + \kappa_{\rm lin}$. \label{figexamplemaps}}
\end{figure*}

 As an example, in the left panels of Fig.~\ref{figexamplemaps} we
 show nine effective linear convergence maps $\kappa_{\rm eff,lin}$
 for the same light-cone realisation as presented in
 Fig.~\ref{figmaps}.  The amplitude of the corresponding power spectra
 has been sampled using a Gaussian random number generator and
 adopting $P_{\rm \kappa, lin}(l)$ as theoretical reference model and
 renormalized according to the relation in eq.~(\ref{eqampl}).  In
 real space, each effective linear map is in phase with the non-linear
 structures present in the field of view and statistically consistent
 with the matter density distribution in sheets, filaments and
 knots. The right panels of the figure show the total effective maps
 summing the maps in the left panels with the convergence maps
 constructed from haloes and subhaloes as in Fig.~\ref{figmaps}.

\begin{figure*}
  \includegraphics[width=0.3\hsize]{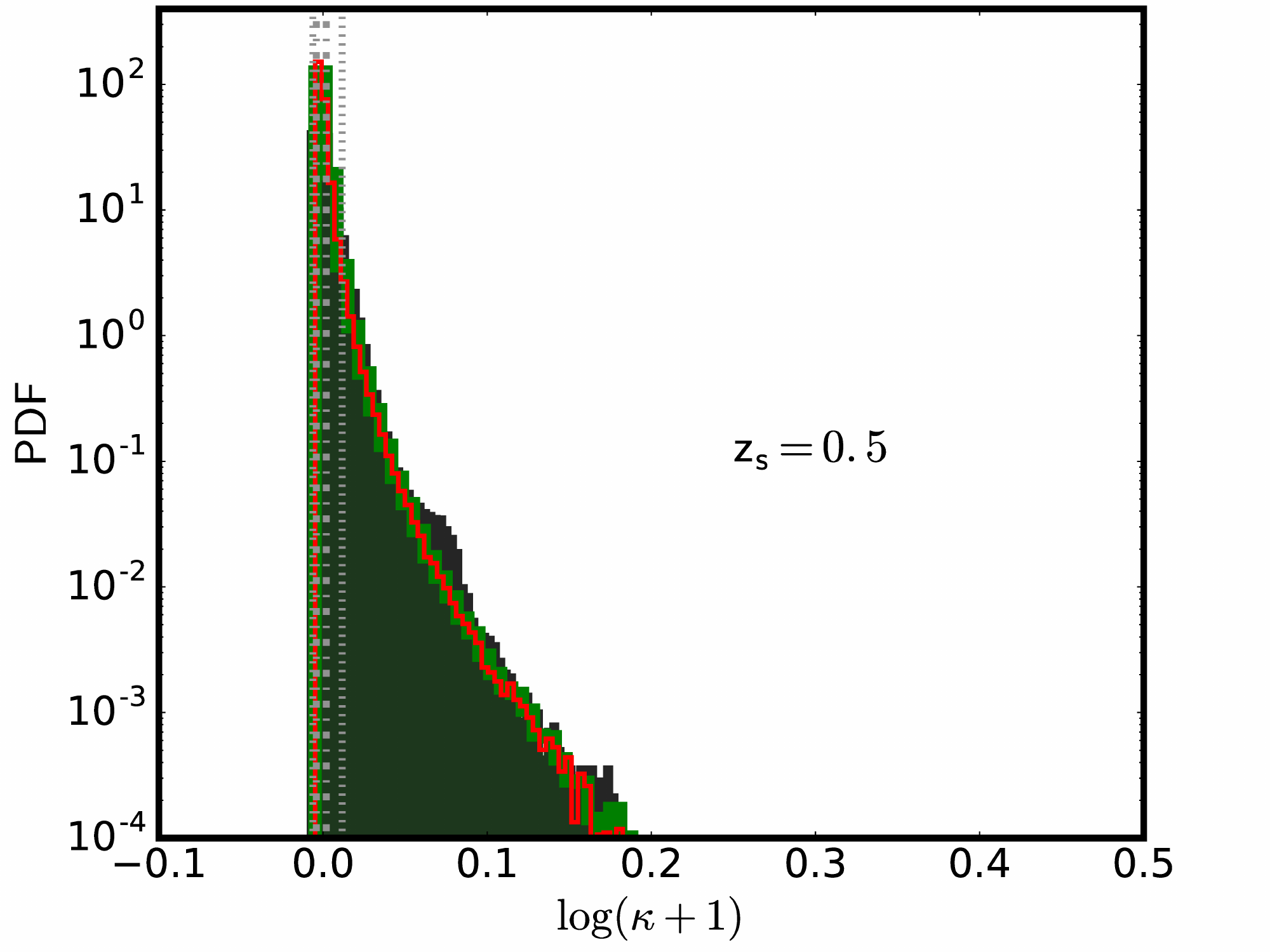}
  \includegraphics[width=0.3\hsize]{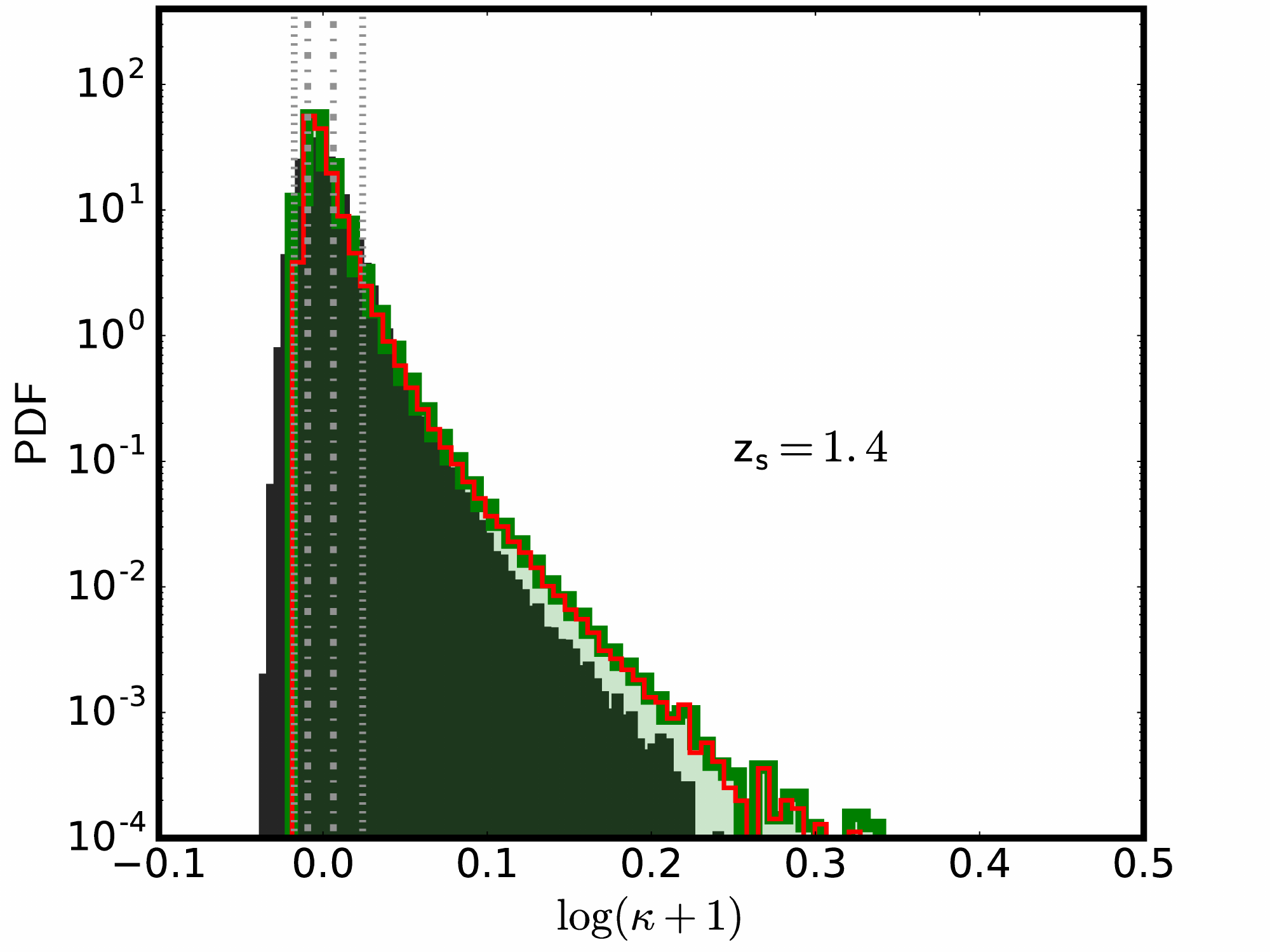}
  \includegraphics[width=0.3\hsize]{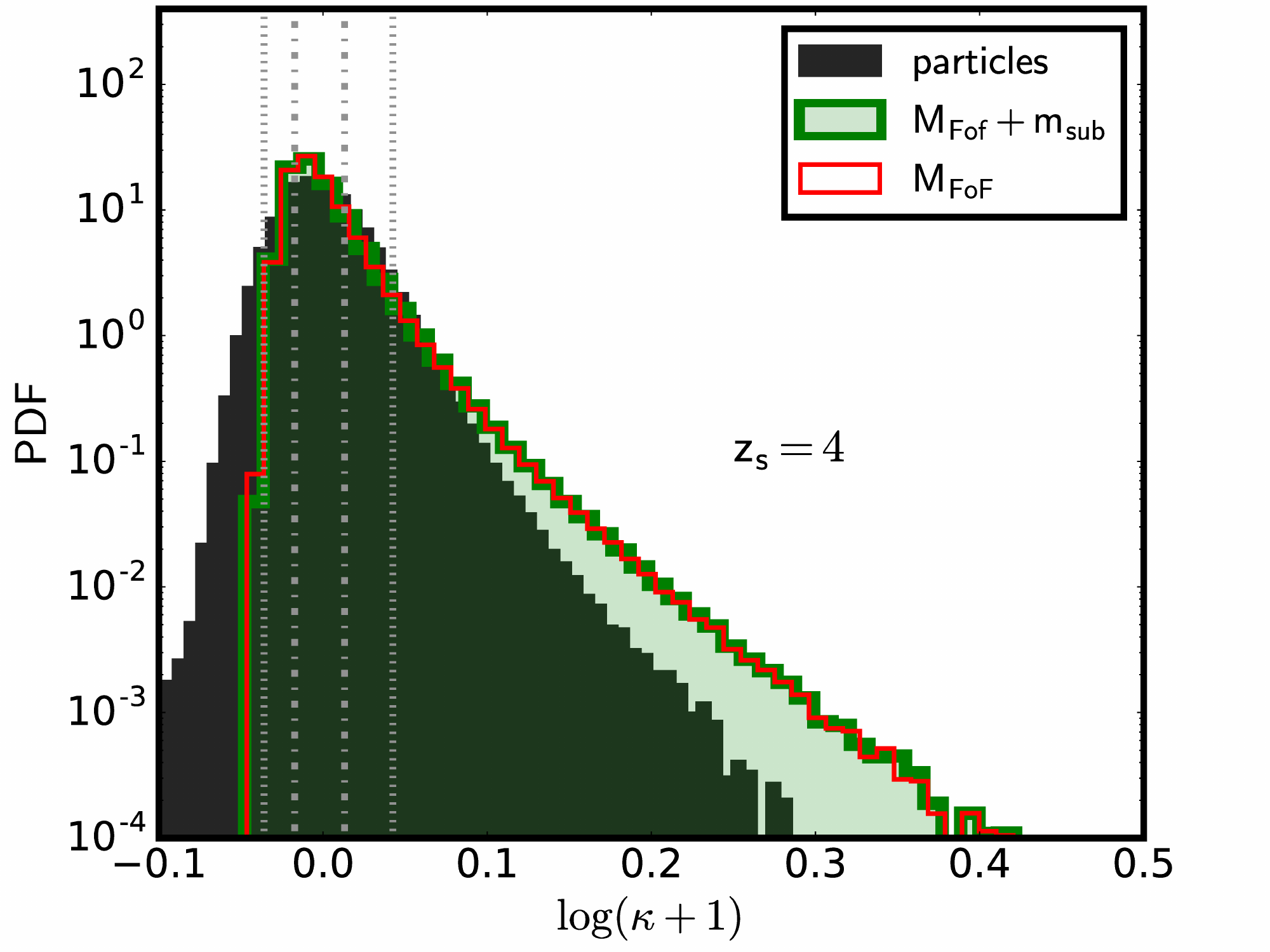}
  \caption{PDF of the convergence maps constructed at three source
    redshifts. As in Fig.~\ref{pdf00fig} the black histograms present
    the distribution of the convergence maps built from the
    ray-tracing pipeline. The green and the red histogram refer to the
    effective convergence maps build from the FoF groups and the FoF
    plus subhaloes including also the effective Gaussian linear
    contribution. Here we show the same light-cone realisation as in
    Fig.~\ref{figmaps} for which we have created $64$ different
    realisations of the Gaussian linear power spectrum -- as an
    example we have displayed nine of them in
    Fig.~\ref{figexamplemaps}.\label{figpdfniosed}}
\end{figure*}  

\begin{figure}
\includegraphics[width=\hsize]{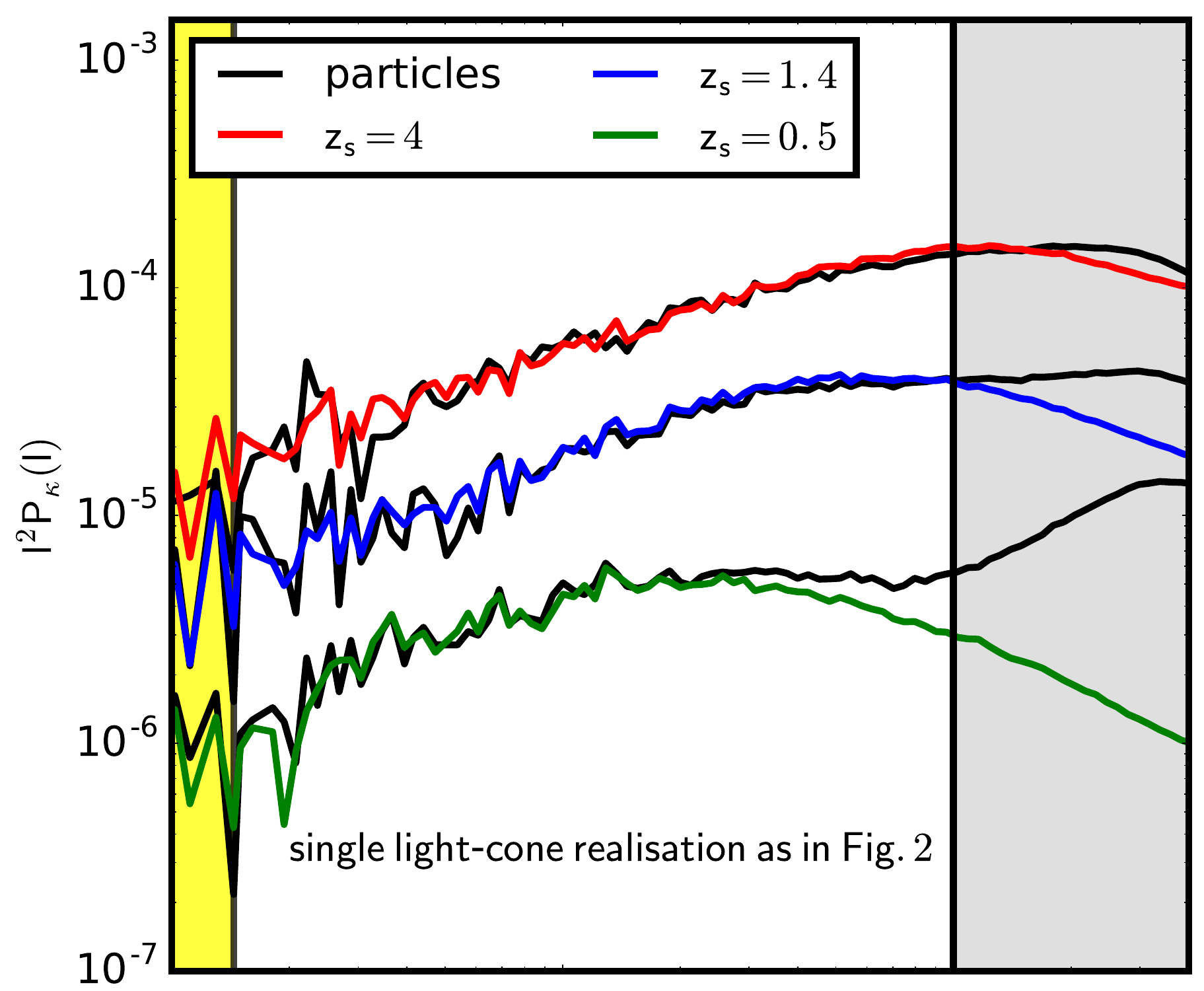}
\includegraphics[width=\hsize]{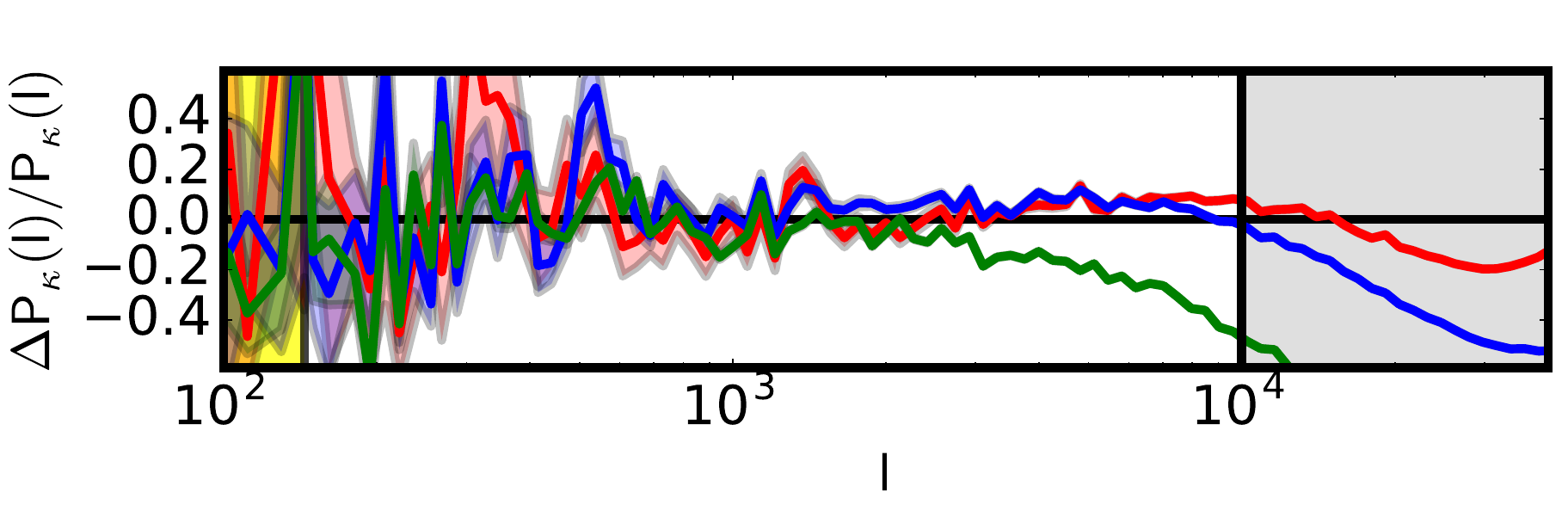}
  \caption{\label{figpowerComp}Top panel: convergence power spectra at
    three different source redshifts of a single light-cone
    realisation. The black curves show the measurement computed from
    the ray-tracing using particles while the green, red and blue
    display the prediction from our halo model algorithm -- including
    FoF haloes and subhaloes -- plus the effective linear contribution
    to resemble the matter density distribution not present in
    haloes. For the effective linear contribution we have generated
    $64$ different random Gaussian maps in amplitude all with the same
    phase and measured the average.  Bottom panel: relative difference
    of the power spectra, the corresponding shaded regions enclose the
    variance of the power spectra on $64$ different random Gaussian
    realisation of the effective linear contribution.\label{figpk1}}
\end{figure}  

\section{Statistical Properties of the \textsc{WL-MOKA\_Halo-Model}}
\label{secstatmoka}
The effective total maps reconstructed using our halo model reproduce
quite well the properties of the maps computed using all the particles
in the simulation that are present within the light-cones up to a
given source redshift $z_s$. The halo and the subhalo catalogues are
used to compute the contributions from non-linear structures while the
linear power spectrum is used to characterise matter not located in
haloes.

\begin{figure}
  \includegraphics[width=\hsize]{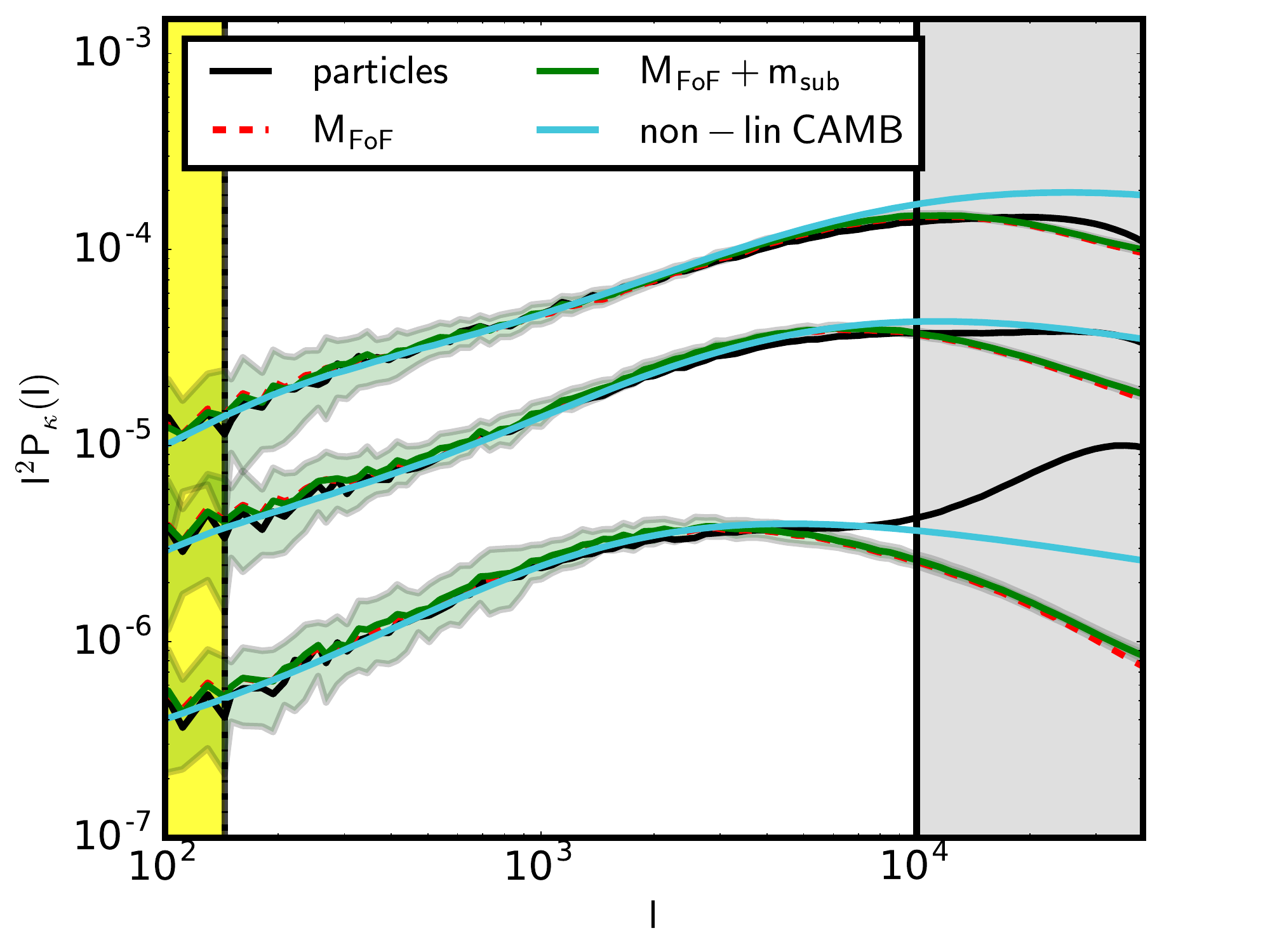}  
  \caption{Reconstructed non-linear convergence power spectrum at
    three different source redshifts: $z_s=4$, $z_s=1.4$ and
    $z_s=0.5$, from top to bottom, respectively. The black solid lines
    represent the prediction using the ray-tracing pipeline on the
    particles within the light-cones. Red dashed lines show the
    results using Friends-of-Friends haloes, while the green solid lines
    show the case of FoF haloes plus subhaloes. Shaded green regions
    enclose the rms of the $25$ light-cones realisations, for each we
    created $64$ different realisations of the effective linear
    Gaussian contribution. The cyan curves exhibit the non-linear
    predictions obtained from CAMB \citep{camb}, which implements the
    \citet{takahashi12} version of HALOFIT
    \citep{smith03}.\label{fignonlinearkappafinal}}
\end{figure}

In Fig.~\ref{figpdfniosed}, we display the PDF of the convergence maps
for the first light-cone realisation comparing the effective total
maps (haloes -- and subhaloes -- plus the effective linear term) with
the maps computed from particles -- as in Fig.~\ref{pdf00fig}.  For
the maps constructed using our \textsc{WL-MOKA\_Halo-Model} (where
\textsc{MOKA} stands for Matter density distributiOn Kode for
gravitationAl lenses) we have generated $64$ random realisations of
the amplitude of the effective linear contribution.  Left, central and
right panels show the results for $z_s=0.5$, $1.4$ and $4$,
respectively. Again we notice that the PDF of the maps constructed
using FoF haloes and subhaloes has a more extended tail toward larger
values of the convergence with respect to the maps from simulation:
this is due to the fact that they resolve much better the centre of
haloes and subhaloes that may suffer from finite mass and force resolution when
using particles. Our halo model runs are only limited by the size of
the $\kappa$ map we set equal to $2048 \times 2048$. This corresponds
to approximately $8.8$ arcsec per pixel.. Comparing the green and the
red histograms we can notice that including subhaloes the maps present
pixels with larger values of the convergence which correspond to the
clump cores within FoF groups. From the figure we can also see that
the distributions for $\kappa<0$ presents a different sampling of the
convergence field: the black histograms are well described by a
lognormal tail. In Fig.~\ref{figpk1} we show the corresponding
convergence power spectra of the same light-cone realisation and
source redshifts. Black lines are the measured quantities from the
convergence maps computed using particles while green, blue and red
curves the corresponding predictions using FoF and subhaloes plus
effective Gaussian linear term. As it can be seen in the bottom panel
convergence power spectra agree within five percents for angular modes
between the Nyquist frequency and $l \approx 10^4$. It is interesting
to notice that for $z_s=0.5$ the particle shot-noise term starts to
dominate already at $l \approx 2 \times 10^3$.

A more detailed comparison between our \textsc{WL-MOKA\_Halo-Model}
and the ray-tracing analysis can be observed in
Fig.~\ref{fignonlinearkappafinal}, where we show the convergence power
spectra at three different source redshifts, from top to bottom
$z_s=4$, $1.4$ and $0.5$, respectively. The black solid curves show
the average results of $25$ light-cone realisations from the
ray-tracing simulations, the dashed red lines the average cosmic shear
power spectrum of our halo model using only the Friends-of-Friends
groups while the green curves show the average using FoF with
subhaloes.
The light-green shaded regions display the rms corresponding to the
average measurement of the \textsc{WL-MOKA\_Halo-Model} including
haloes and subhaloes.  The cyan curves are the predictions from CAMB
using the prescription of \citet{takahashi12} for the non-linear
modeling.  We would like to underline that possible small departures
at small angular modes between our \textsc{WL-MOKA\_Halo-Model}
predictions and the results from the ray-tracing simulation may be due
to the fact that while we generally produce a large sample of Gaussian
random realisations of the linear theoretical predictions, in the
simulation we have only one random realisation of the initial density
field as computed at $z_i=99$.

Using cosmic shear measurements to estimate cosmological parameters
requires a good knowledge of the covariance matrix, this means
information about the correlation and cross-correlation of the lensing
measurements between different angular scales, or modes.  Typically to
have a good sampling of the covariance matrix, we need thousands of
independent light-cone realisations of the same field of view and for
different initial conditions for the same cosmological model
\citep{taylor14}. Using numerical simulations and full ray-tracing
analyses, the production of these light-cones requires an enormous
amount of computational time and huge storage disk spaces. On the
other hand our approximated halo model approach is much faster, not
too much demanding in terms of CPU time and little memory consuming,
and only requires as input the halo and subhalo catalogues present
within the field of view up to the required source redshift plus the
linear power spectrum. 

\begin{figure*}
  \includegraphics[width=0.32\hsize]{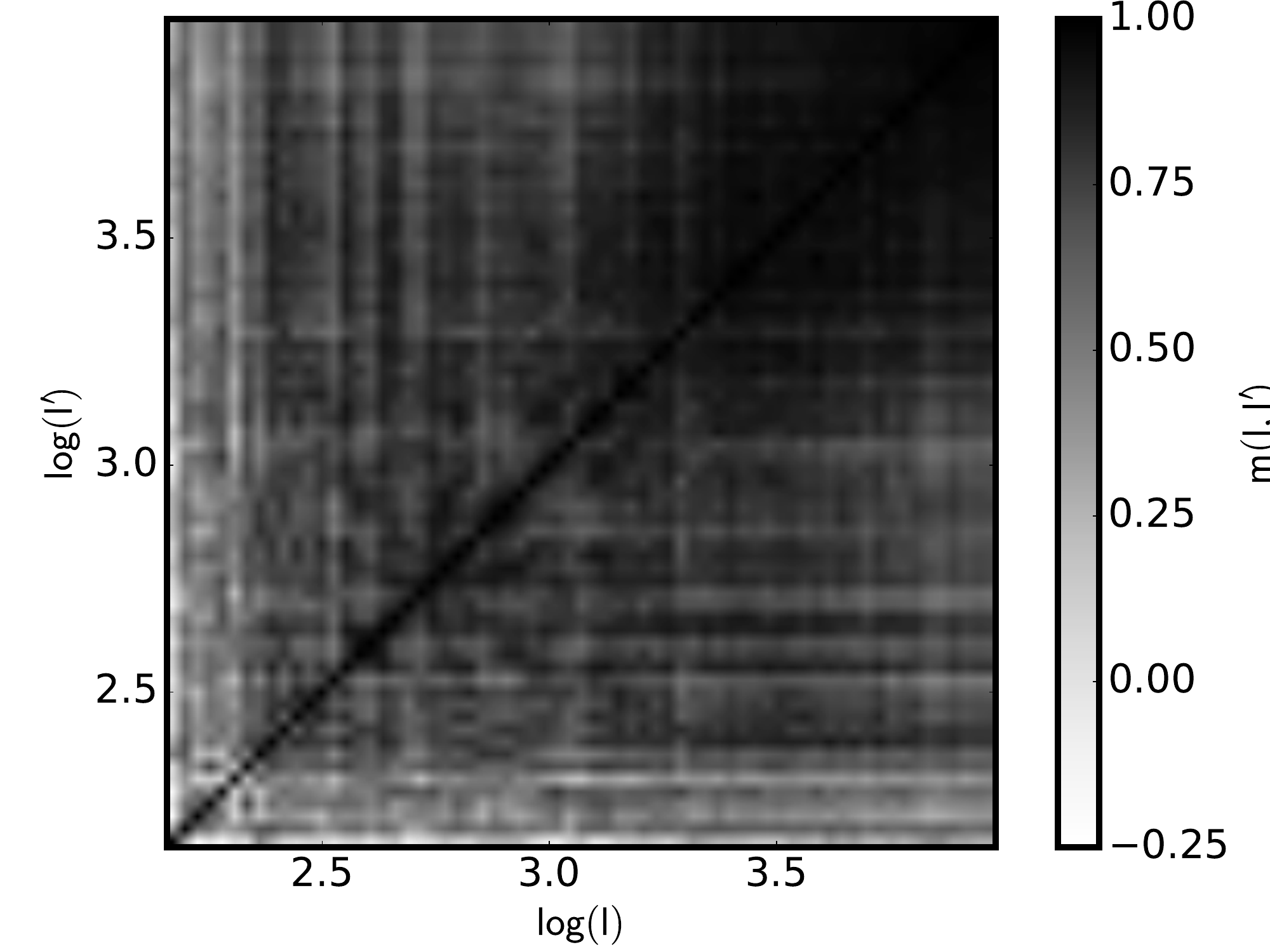}  
  \includegraphics[width=0.32\hsize]{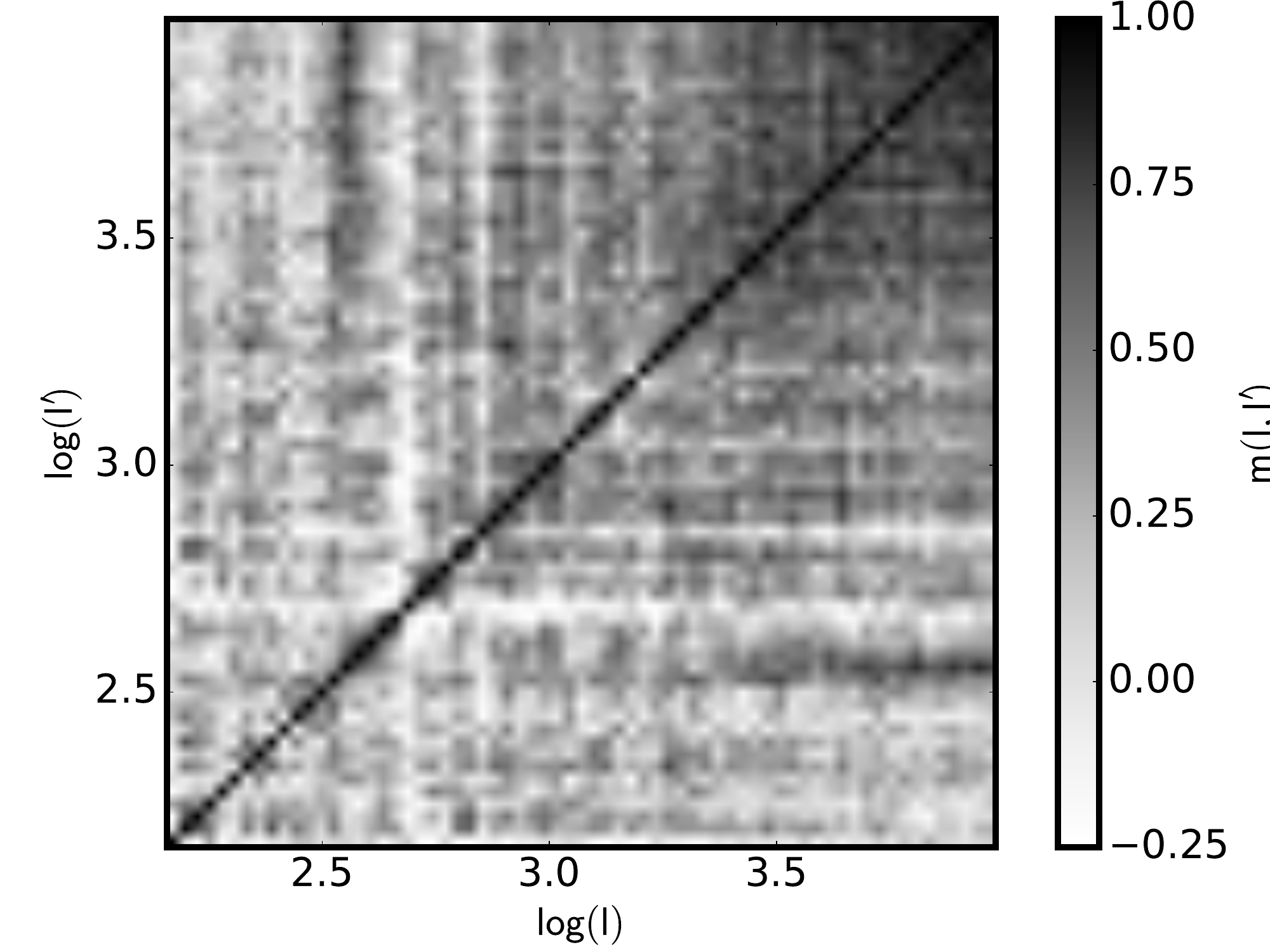}
  \includegraphics[width=0.32\hsize]{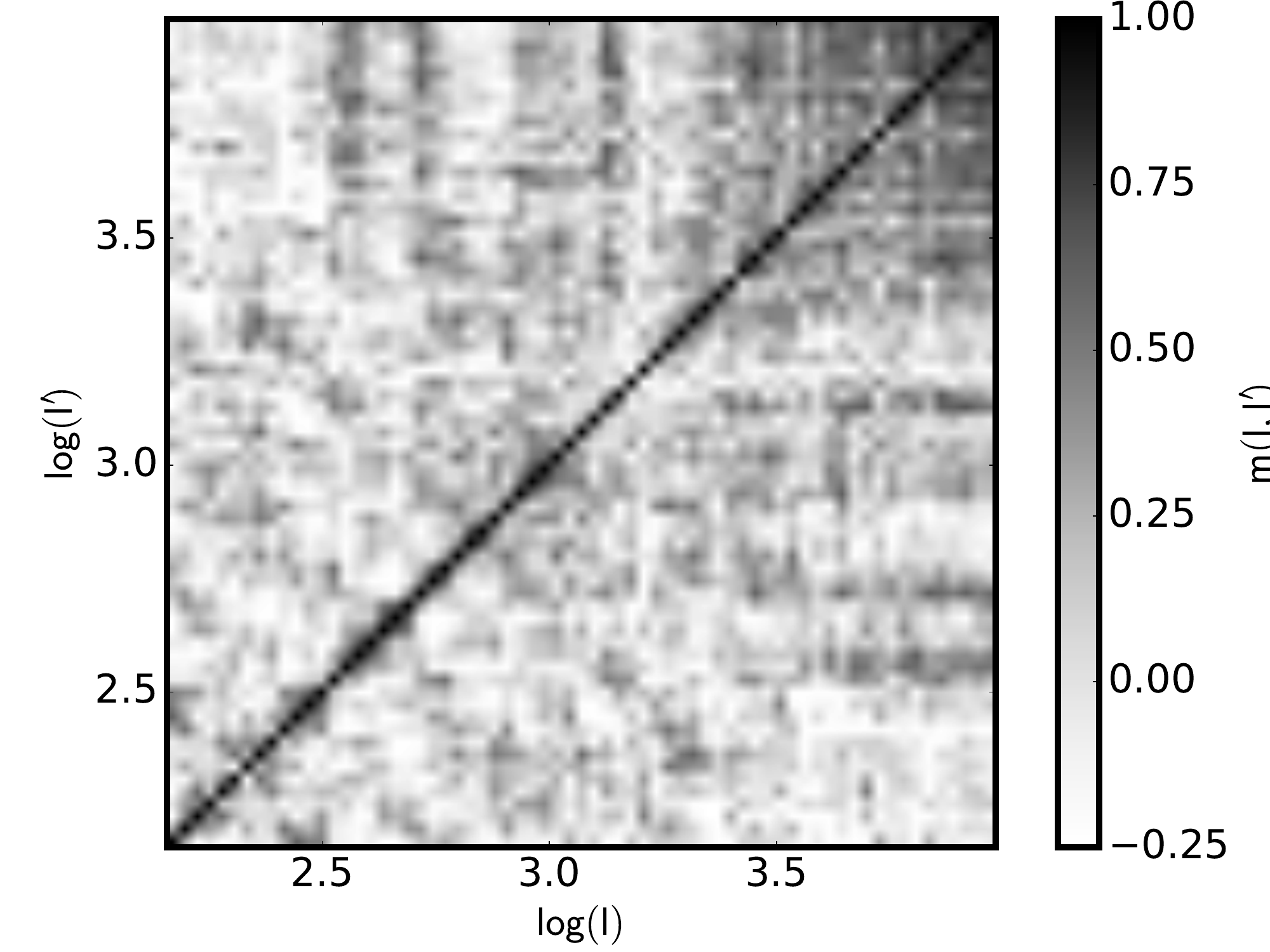}    
  \includegraphics[width=0.32\hsize]{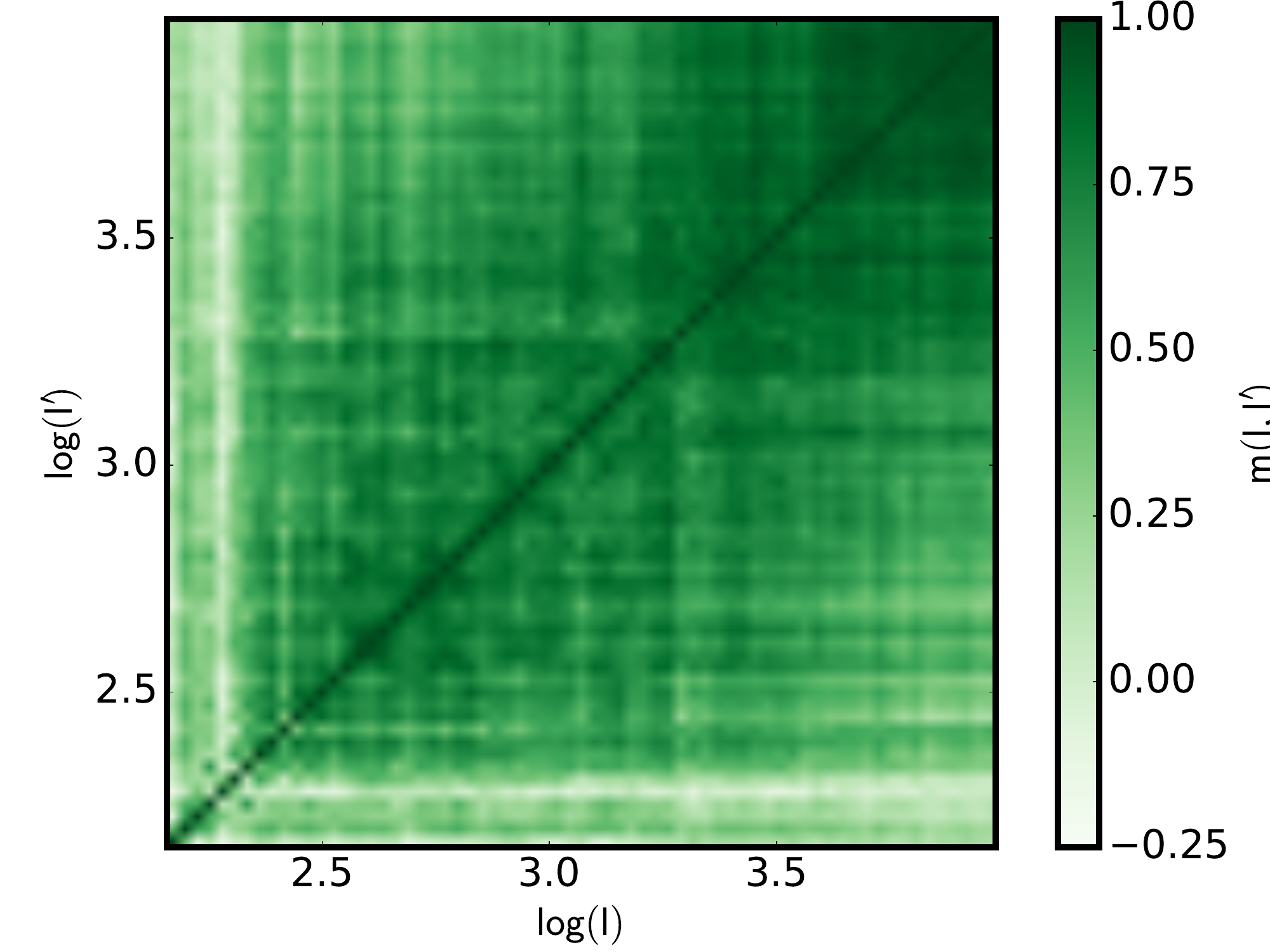}
  \includegraphics[width=0.32\hsize]{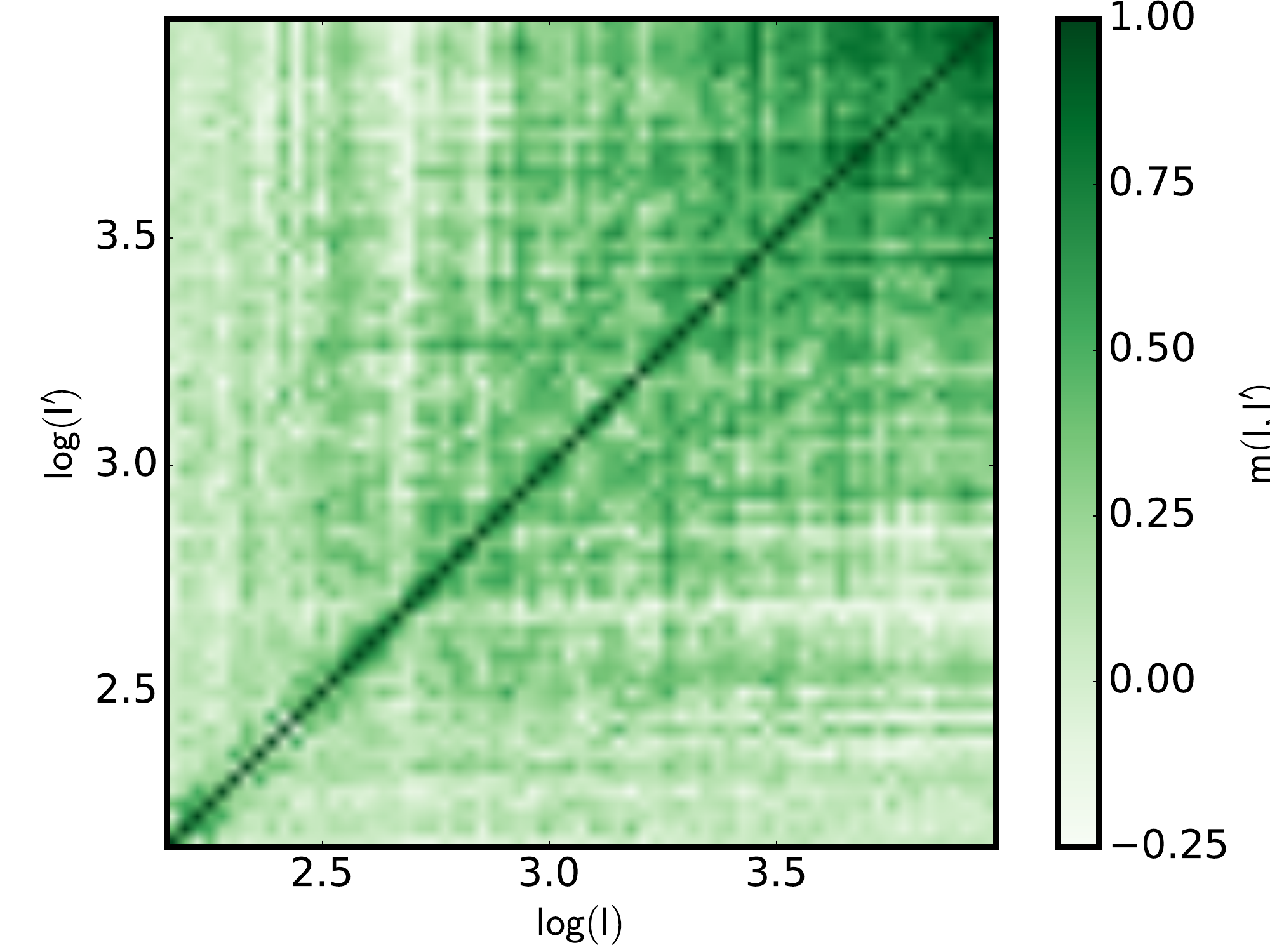}
  \includegraphics[width=0.32\hsize]{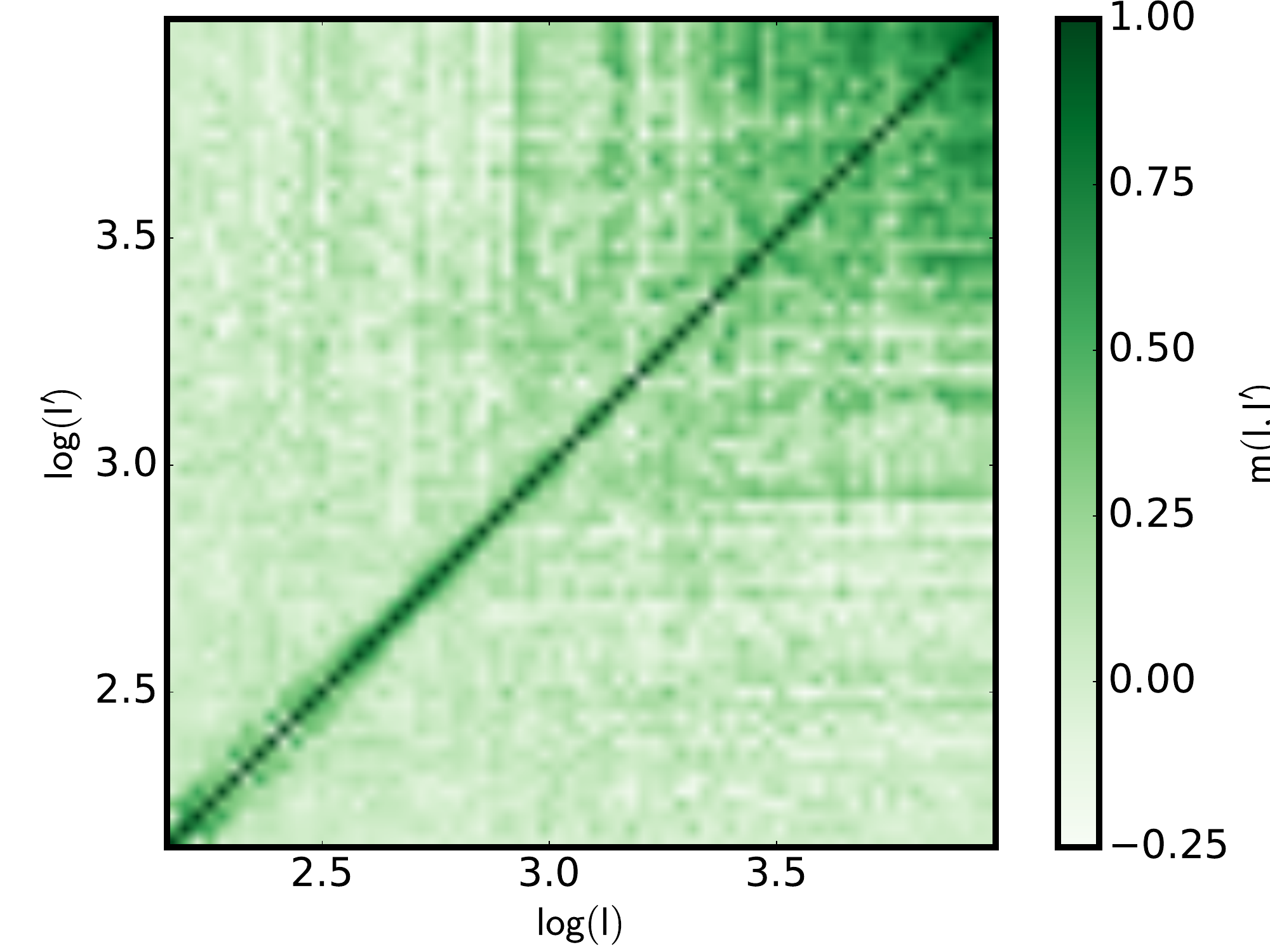}
  \caption{Correlation matrices at three different source
    redshifts. From left to right $z_s=0.5$, $z_s=1.4$ and $z_s=4$. On
    the first row, the panels show the correlation matrices of $25$
    different light-cone realisations of the ray-tracing pipeline
    while in the second row are displayed the matrices from the FoF
    haloes populated with subhaloes plus the effective Gaussian linear
    contribution. In this case for each light-cone created using the
    halo model realisation we have generated $64$ random Gaussian
    effective maps to account for the matter density distribution that
    is not in haloes.\label{figcov}}
\end{figure*}

From the different light-cones realizations
we can write the covariance matrix in Fourier space as:
\begin{equation}
M(l,l') = \langle P_{\kappa}(l) - \bar{P}_{\kappa}(l)\rangle\langle P_{\kappa}(l') - \bar{P}_{\kappa}(l')\rangle
\end{equation}
where $\langle \bar{P}_{\kappa}(l)\rangle$ represents the best
estimate of the power spectrum at the mode $l$ obtained from the
average of all the corresponding light-cone realisations and
$P_{\kappa}(l)$ represents the measurement of one realisation.  The
matrix can be then normalised as follows to obtain the correlation matrix
\begin{equation}
m(l,l') = \dfrac{M(l,l')}{\sqrt{M(l,l) M(l',l')}}\,.
\end{equation}
For comparison with the $25$ independent light-cones generated from
the ray-tracing simulation using particles in the top panels of
Fig.~\ref{figcov} we show the correlation matrices of the cosmic shear
power spectrum of those different realisations assuming $z_s=0.5$,
$1.4$ and $4$, from left to right, respectively.  We remind the reader
that this is presented here only for comparison. On the second row of
the figure (in green scale), the three panels show the correlation
matrices computed using our \textsc{WL-MOKA\_Halo-Model} formalism. We
have computed the halo and subhalo contributions from the $25$
different light-cone realisations and for each of them we have
generated $64$ effective linear term contribution to represent the
matter density distribution that is not in haloes. From the figure, we
notice that our halo model reconstructs with very good accuracy the
halo sampling properties of the non-linear structures and the
contribution from linear theory typically dominant for small values of
$l$ within the field of view realization. This sets the
basis for the capability of our approach to create self-consistent
covariance matrices that can be easily extended to much larger field
of view, accounting for a uniform or masked fields of view and
considering different geometries and determining how these properties propagate into the lensing measurements and subsequently
into the covariance matrices \citep{harnois-deraps15b}.

Before concluding this section we would like to discuss the
performance of our halo model-based weak lensing methods in comparison
to the full ray-tracing simulation using particles. The first bottle
neck in making convergence maps using particles is the construction of
the lensing planes and reading the simulation snapshot
files. Typically for a $1024^3$ dark matter particle simulation the
construction of a plane resolving a field of view of $5 \times 5$
sq. degrees with $2048^2$ pixels takes $2.5$ min that for 22 lens
planes up to redshift $z=4$ translates in approximately $60$
min. While the construction of the corresponding halo and subhalo
catalogues, reading and projecting the \textsc{subfind} catalogues
within the same field of view, takes slightly less then $1.5$ min. The
full ray-tracing simulation with \textsc{glamer} on $22$, $14$ and $8$
lens planes, which are needed to construct the convergence maps and
measure the convergence power spectrum at $z_s=4$, $z_s=1.4$ and
$z_s=0.5$, consumes $70$, $65$ and $62$ min ($8$ threads process),
respectively while our halo model code (single thread process) takes
$75$ min on haloes in a $5 \times 5$ sq. degrees.  This time almost
doubles when we want to account also for a buffer region of $2.5$
degrees on a side. On a single light cone simulation our fast halo
model method is approximately $90\%$ faster than the full ray-tracing
simulation using particles. However, it should be stressed that a
N-body run from $z=99$ to the present time using the \textsc{Gadget2}
code \citep{springel05a} takes around $50.000$ CPU hours, while a run
with an approximate method like \textsc{Pinocchio}\footnote{In
  particular a run at galileo@cineca (32 core) $1024^3$ takes $15$
  min.}  \citep{monaco13} takes approximately $750$ hours to generate
also the past light-cone up to the desired maximum redshift $z=4$ with
our same aperture using a $512^3$ grid -- on which we can run our fast
weak lensing method -- while it spends $1550$ CPU hours for the same
simulation but using a finer grid of $1024^3$ \footnote{All the CPU
  times given here have been computed and tested in a 2.3 GHz
  workstation.}.  To summarise, we notice that our fast weak lensing
simulation plus an approximate N-body method for the halo catalogue
are much faster than the full-ray tracing simulation plus an N-body
solver, but still reaching the same level of accuracy in the
convergence power spectrum.

\section{Summary \& Conclusions}
\label{sumandcon}
In this paper we have presented a self-consistent halo model formalism
to construct convergence maps with statistical properties compatible
with those derived from the full ray-tracing pipeline.

From the $\Lambda$CDM run of the CoDECS suite we have produced
catalogues of haloes and subhaloes present within the constructed
matter density light-cones of a field of view of $5 \times 5$
sq. degrees up to $z_s=4$. To avoid border effects, we stored the
information about the haloes and the subhaloes present in a field of
view of $10 \times 10$ sq. degrees. In the following points we
summarise the main ingredients and results of our analyses:
\begin{itemize}
\item the mass density distribution in haloes is modelled using the
  NFW profile. For concentration, we adopt the model by \citet{zhao09}
  for the Friends-of-Friends groups while the \citet{giocoli13}
  function for the $M_{\rm 200}$ mass definition;
\item the positive part of the one point statistic of the convergence
  field is quite well reconstructed using the halo model formalism,
  however using only the matter present in haloes and subhaloes we
  are missing the linear matter density field not attached to
  virialized structures -- this means in particular filaments and
  sheets of the cosmic web;
\item the power spectrum of the density field reconstructed with
  haloes reflects the absence of matter outside haloes, and present
  less power at large scale than as expected from linear theory;
\item the subhalo contribution, using truncated Singular Isothermal
  Sphere profile, enhances the convergence power spectrum by
  approximately $3\%$ up to $l \approx 10^4$. At smaller scales, this
  contribution increases dramatically;
\item the effective linear contribution on large scales is included by
  creating a Gaussian field from the theoretical linear cosmic shear
  power spectrum coherent in phase with the distribution of haloes
  present in the simulated field of view, renormalizing it in
  amplitude in order to match the linear prediction on large scales;
\item the total effective maps are statistically similar to the
  ray-tracing ones constructed using the particle density field.
\end{itemize}

To summarise, our \textsc{WL-MOKA\_Halo-Model} formalism
self-consistently reconstructs the statistical properties of matter
density distribution within light-cones only using the halo and
subhalo properties plus the linear power spectrum of the considered
cosmological model. When compared with a full ray-tracing simulation
using particles for each single realisation, we find an agreement on
average within $5\%$ with the reconstructed convergence power spectra
for different source redshifts.  This highlights the capability of our
halo model pipeline in reconstructing the non-linear properties of
weak-lensing fields in a much faster way than ray-tracing
simulations. Future tests will be dedicated to the capability of
extend our method to non-standard cosmologies (Giocoli et al. in
preparation) in the light of the recent results presented by
\citep{narikawa11,zhang13,massara14,lombriser15,mead16} and also
to the possibility to self-consistently develop general models for the
cross-correlation between clustering and weak-lensing signals
\citep{delatorre16}.

Our formalism opens the capacity to create coherent
covariance matrices for a given cosmological model and any field of
view geometry and masking, allowing a more complete and
self-consistent cosmological inspection of realistic lensing data over
a wider range of cosmological parameters
\citep{kids,des,lsst,euclidredbook}.

\section*{Acknowledgments}                                               
CG thanks CNES for financial support.  CG  and MB
acknowledge support from the Italian Ministry for Education,
University and Research (MIUR) through the SIR individual grant
SIMCODE, project number RBSI14P4IH.  EJ and SdlT acknowledge the
support of the OCEVU Labex (ANR-11-LABX-0060) and the A$^\star$MIDEX
project (ANR-11-IDEX- 0001-02) funded by the "Investissements
d'Avenir" French government program managed by the ANR. LM thanks the
support from PRIN MIUR 2015 ``Cosmology and Fundamental Physics:
Illuminating the Dark Universe with Euclid''. LM acknowledges the
grants ASI n.I/023/12/0 Attivit\'a relative alla fase B2/C per la
missione Euclid.  RBM's research was partly part of project GLENCO,
funded under the European Seventh Framework Programme, Ideas, Grant
Agreement n. 259349.  GC acknowledges the organizers of the Light-Cone
and of the Simulation meetings in Garching and Barcelona, particularly
Carmelita Carbone for useful discussions. We thank also Pierluigi
Monaco for reading one of the first version of our manuscript.  CG is
grateful also the Ravi K. Sheth for his hospitality at UPENN and
useful discussions about the idea of this work.

\appendix

\section{Probability Distribution Function of the convergence maps}
As discussed in the text, and more in particular displayed in
Fig.~\ref{figpdfniosed}, the comparison of Probability Distribution
Function (PDF) between our \textsc{WL-MOKA\_Halo-Model} predictions
and those using ray-tracing with particles shows some difference that
varies as a function of the source redshift. In the discussion we have
stressed that this may be due to numerical resolution limits both in
force and particle mass that do not allow for resolving well the
central part of the haloes and clumps where typically high convergence
values appear. However different authors
\citep{taruya02,hilbert11,clerkin16,patton16,xavier16} have discussed
      that the properties of the convergence one point
      statistic may be characterized by a Gaussian or lognormal
      distribution. \citet{das06} have discussed that small
      perturbations with resolution of $\theta\sim 10$ arcsec and
      $z_s=1$ account for most of the strong lensing cases and that
      the PDF is far superior to the Gaussian or the lognormal. They
      also emphasize that for $z_s=4$ about $12\%$ of the
      strong-lensing cases will result from the contribution of a
      secondary clump of matter along the line of sight, introducing a
      systematic error in the determination of the surface density of
      clusters, typically overestimating it by about some percents.

\begin{figure*}
\includegraphics[width=\hsize]{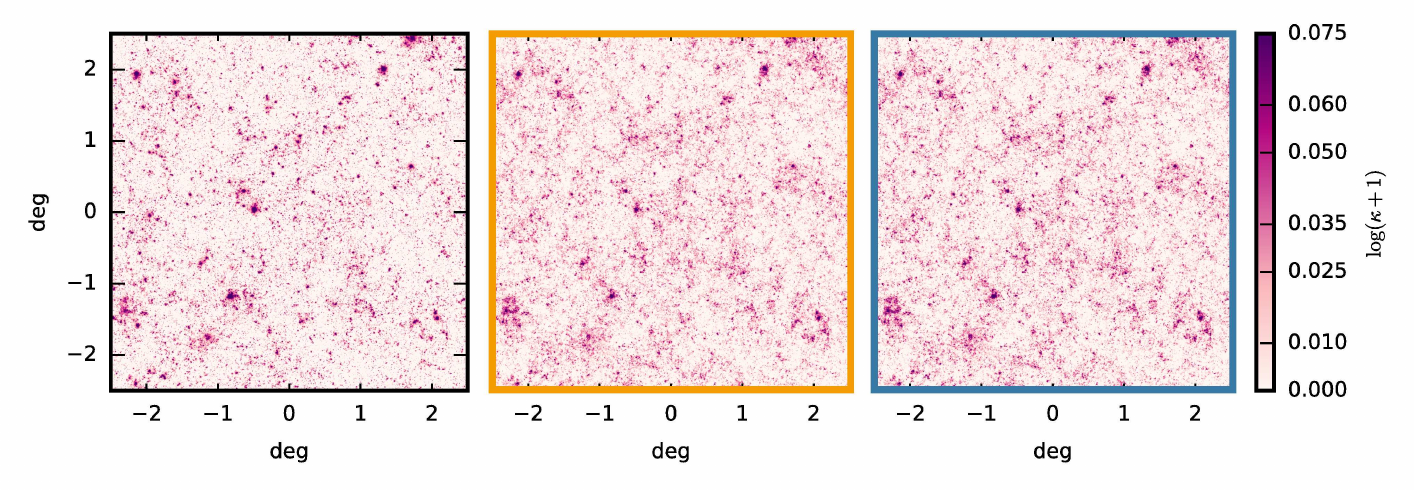}
\caption{Convergence maps of the light-cone constructed considering
  sources located at $z_s=4$. While the left panel displays the
  convergence map produced using our \textsc{WL-MOKA\_Halo-Model}
  algorithm, central and right panel show the same realization of the
  structures with equal phases but forced to have the modulus of the
  convergence field in the Fourier space $\tilde{\kappa}$ randomly
  drawn from a Gaussian (central orange framed) and a lognormal (right
  blue framed) distribution with an identical power spectrum.\label{figmapsapp}}
\end{figure*}

In this appendix we discuss the properties of the PDF of the
convergence resampling the characteristics of the reconstructed fields
in order to have a well defined distribution for the amplitude in the
Fourier space $\tilde{\kappa}$ and conserving both the power spectra
and the phases to be consistent with non-linear structures.  In the
left panel of Fig.~\ref{figmapsapp} we display the convergence map
reconstructed up to source redshift $z_s=4$ using our
\textsc{WL-MOKA\_Halo-Model} algorithm, the map contains the
contributions from haloes, subhaloes and effective linear power
spectrum. The central and right panels show two maps that possess the
same power spectra and coherent in phase with the left one. However,
while in the first (orange framed, termed $resampled_1$) the amplitude
of the convergence in the Fourier space $\tilde{\kappa}(l)$ is drawn
from a Gaussian distribution with rms $\sigma(l)$, in the second (blue
framed, termed $resampled_2$) the amplitude of $\ln(\tilde{\kappa}+1)$
is drawn from a Gaussian distribution with the rms that can be read
as:
\begin{equation}
\sigma^2_{\ln}(l) =  \ln( \sigma^2(l) + 1 )\,
\end{equation}
where $\sigma^2(l) = P_{\kappa}(l)$ and $P_{\kappa}(l)$ the
convergence power spectrum of the map on the left panel. We then
convert the logarithm of the convergence plus one
$\widetilde{\ln(\kappa+1)}$ field in the real space and obtain the
convergence as
\begin{equation}
  \kappa = \exp \left[ \ln(\kappa+1) \right]- 1\,.
\end{equation}
we emphasize that this transformation does generate by construction a
lognormal field in real space \citep{hilbert11,xavier16}, and we
present this case since it produces in real space a map whose PDF is
close to the PDF of the case $M_{\rm FoF}+m_{\rm sub}$.

\begin{figure*}
\includegraphics[width=0.3\hsize]{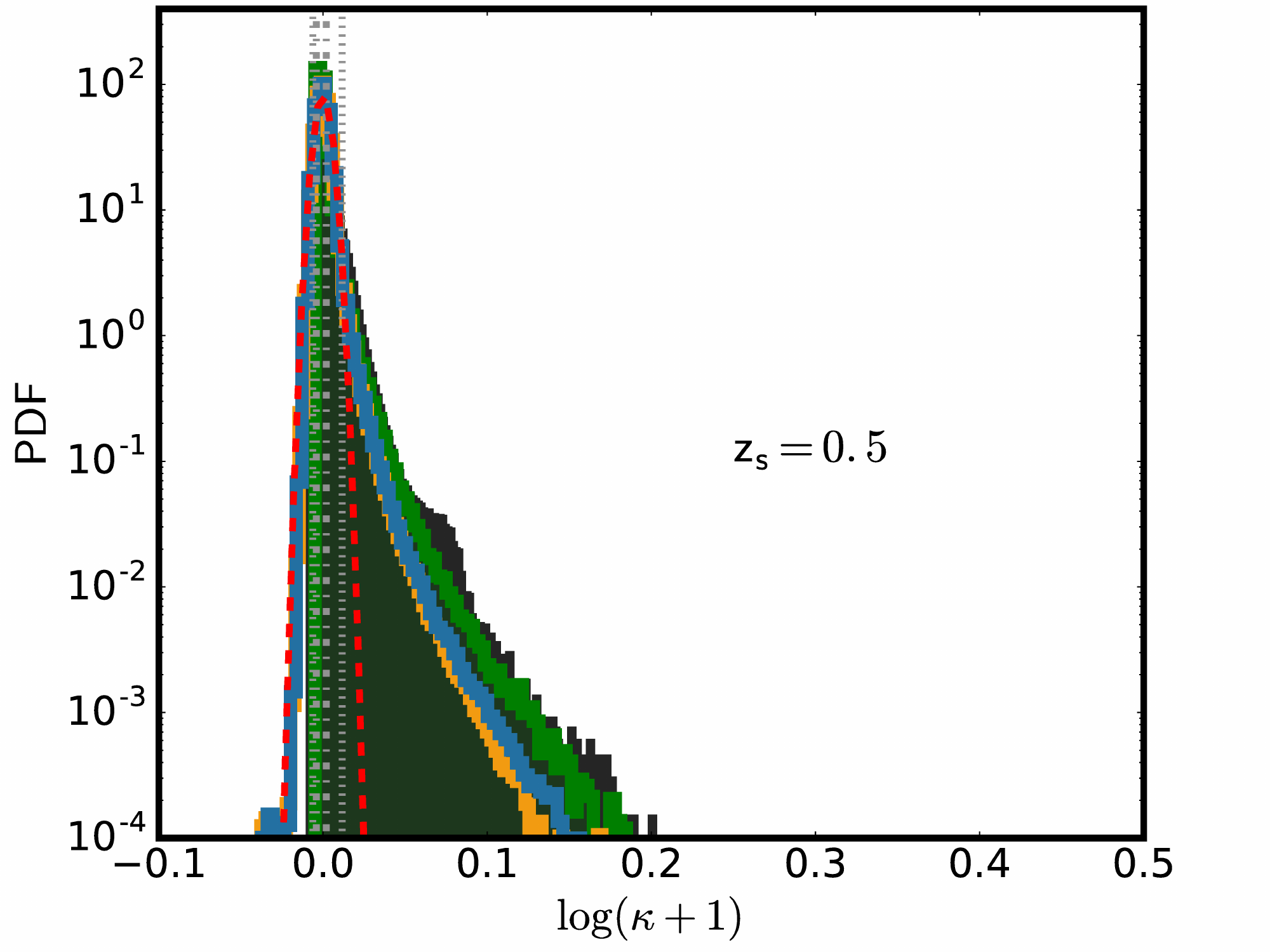}
\includegraphics[width=0.3\hsize]{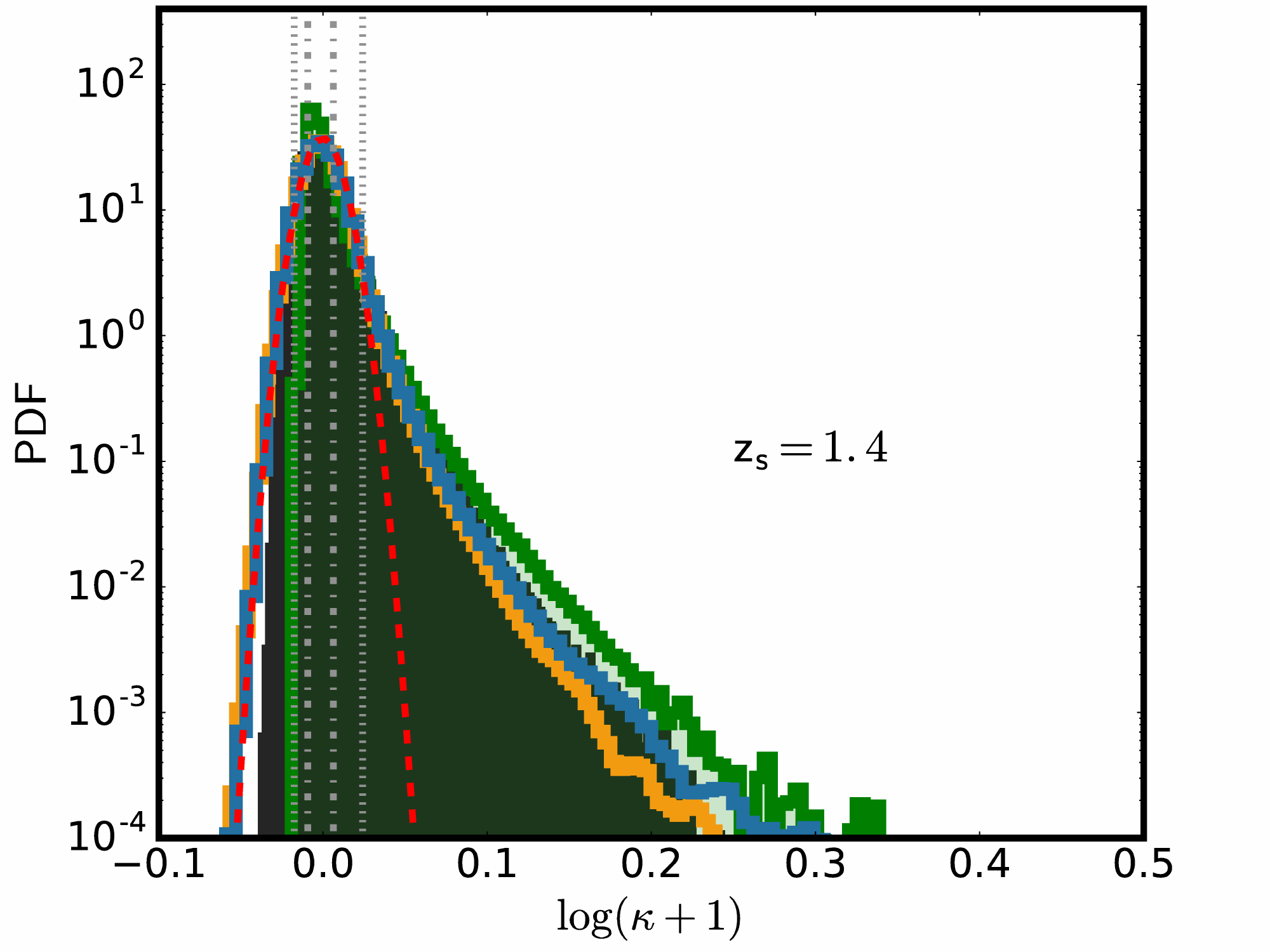}
\includegraphics[width=0.3\hsize]{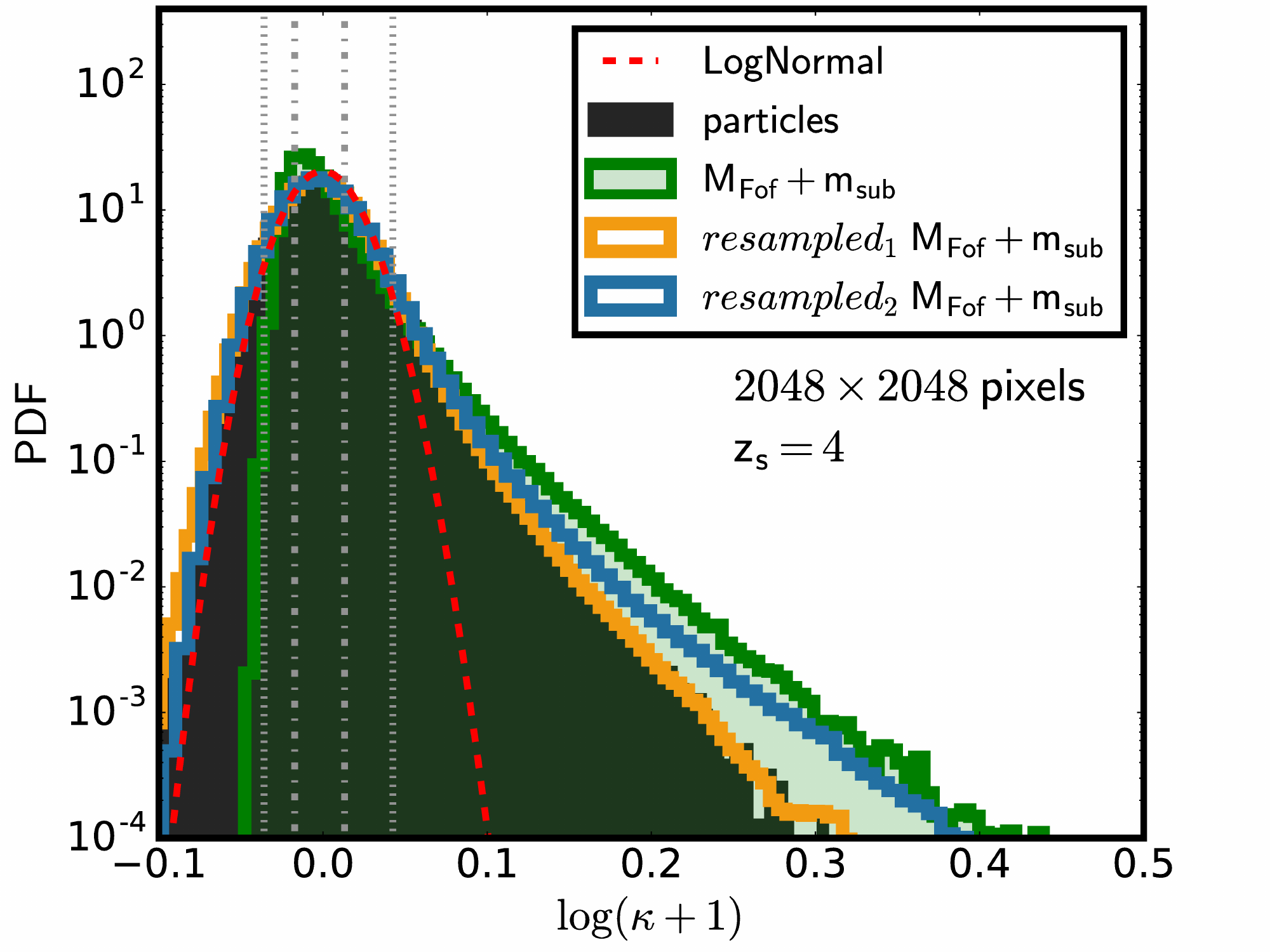}
\includegraphics[width=0.3\hsize]{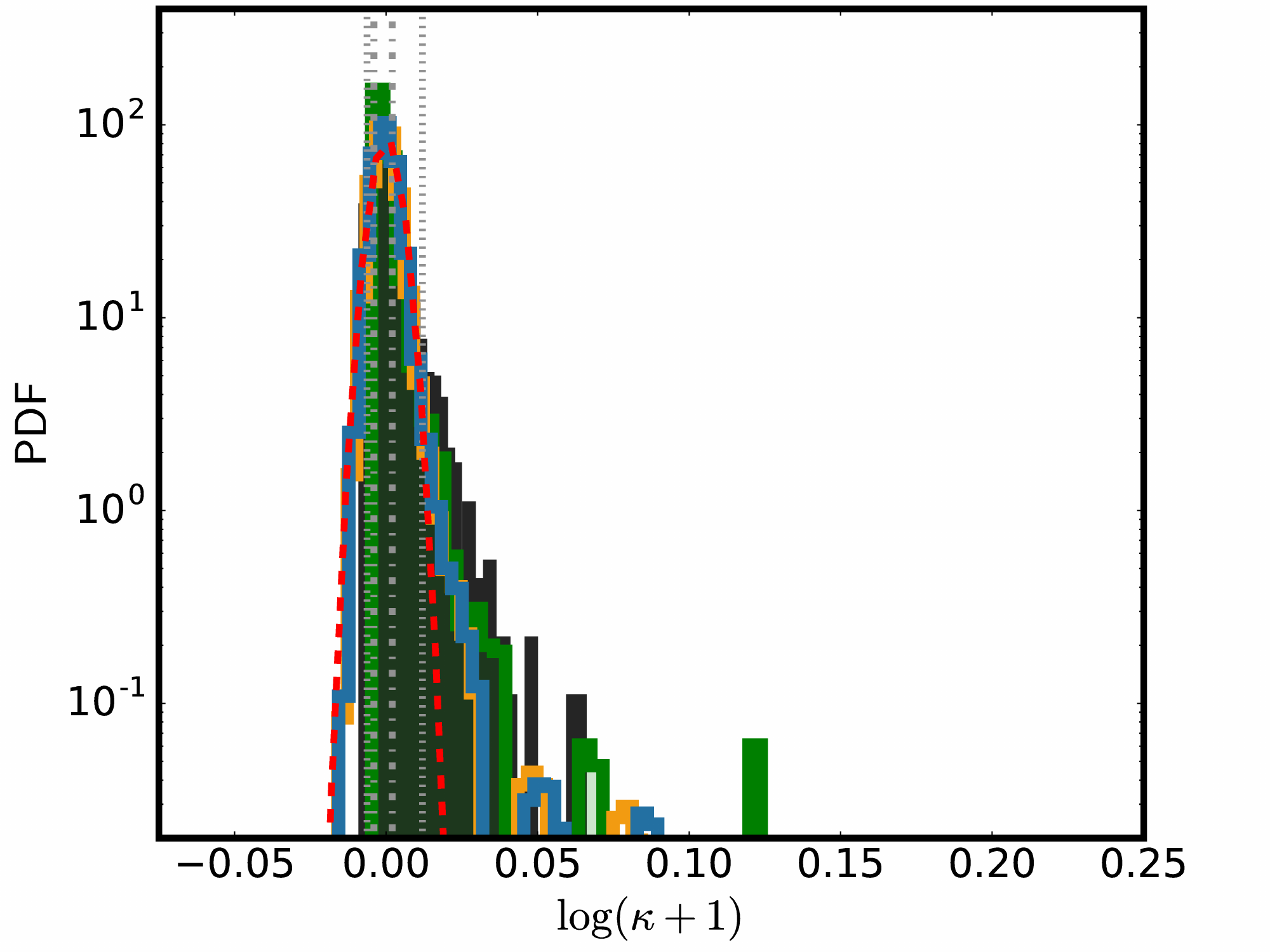}
\includegraphics[width=0.3\hsize]{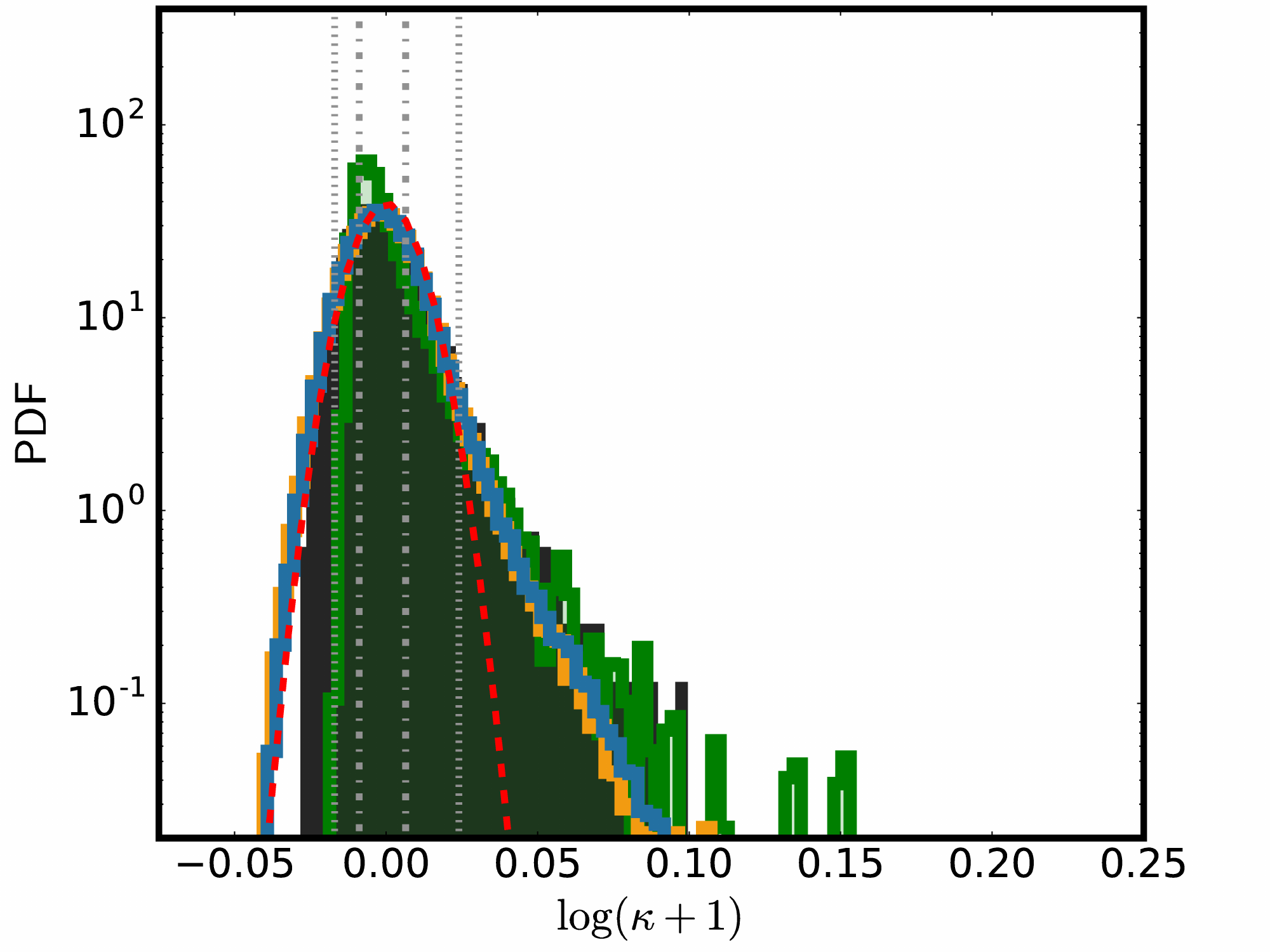}
\includegraphics[width=0.3\hsize]{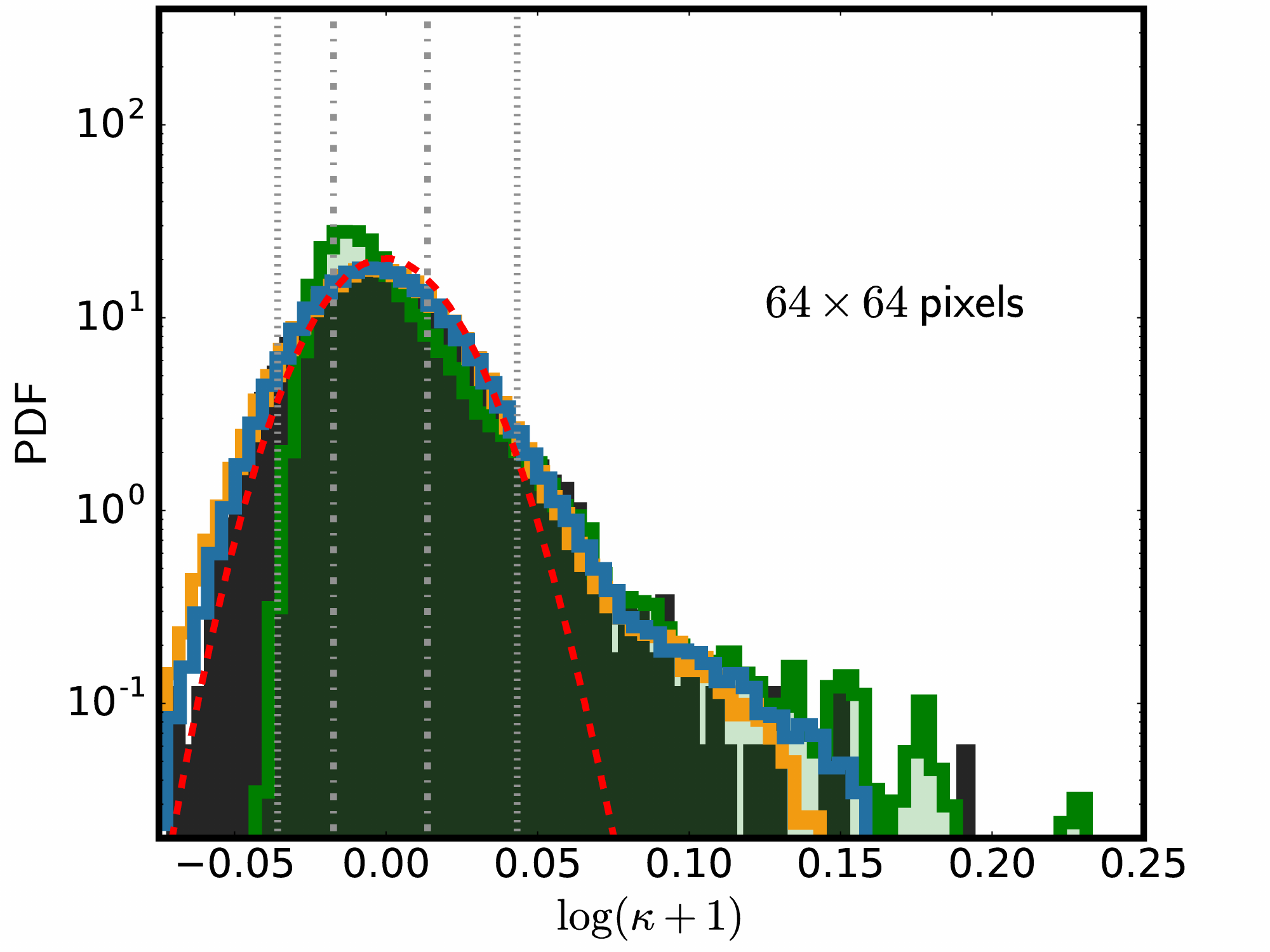}
\caption{Probability Distribution Function of the convergence field
  for the three considered source redshifts, $z_s=0.5$, $1.4$ and $4$
  from left to right, respectively. Top and bottom panels show the PDF
  of the map resolved with $2048$ and $64$ pixels by side,
  respectively. In the top panel the pixel size has a resolution of
  $8.8$ arcsec while in the bottom $281$ arcsec, which correspond to a
  angular mode of approximately $1.5 \times 10^5$ and $4.6\times
  10^3$, respectively.\label{pdffigapp}}
\end{figure*}

In the three top panels of Fig.~\ref{pdffigapp} we exhibit the PDF of
the convergence fields for three different source redshifts, as
labelled in the panels. The black histograms show the PDF of the
convergence field computed using particles and the \textsc{glamer}
pipeline while the green ones the PDF of $64$ realization of the same
field using \textsc{WL-MOKA\_Halo-Model}: haloes, subhaloes and
effective linear power power spectrum contributions. The orange and
blue histograms show the Probability Distribution Function of the
convergence maps resampled in amplitude in the Fourier space as
described above. From the figures we notice that while for low source
redshifts the predictions from numerical simulation are quite close to
the blue histograms for $z_s=4$ the black shaded histogram is very
well described by the orange one.

In the three bottom panels we degrade the resolution of the maps to
$64 \times 64$ pixels which correspond to approximately $281$ arcsec
($l \approx 4.6\times 10^3$) in order to remove the particle noise
contributions. In all panels the red dashed curves show a lognormal
distribution with amplitude equal to half of the first quartile of the
black histograms. In those low resolution maps the one point
distribution function of the convergence is quite well sampled by the
orange histogram, the field is characterized in the Fourier space to
have a Gaussian distribution with average zero and variance at a given
scale given by the square-root of the predicted convergence power
spectrum by our model.

\begin{figure}
\includegraphics[width=\hsize]{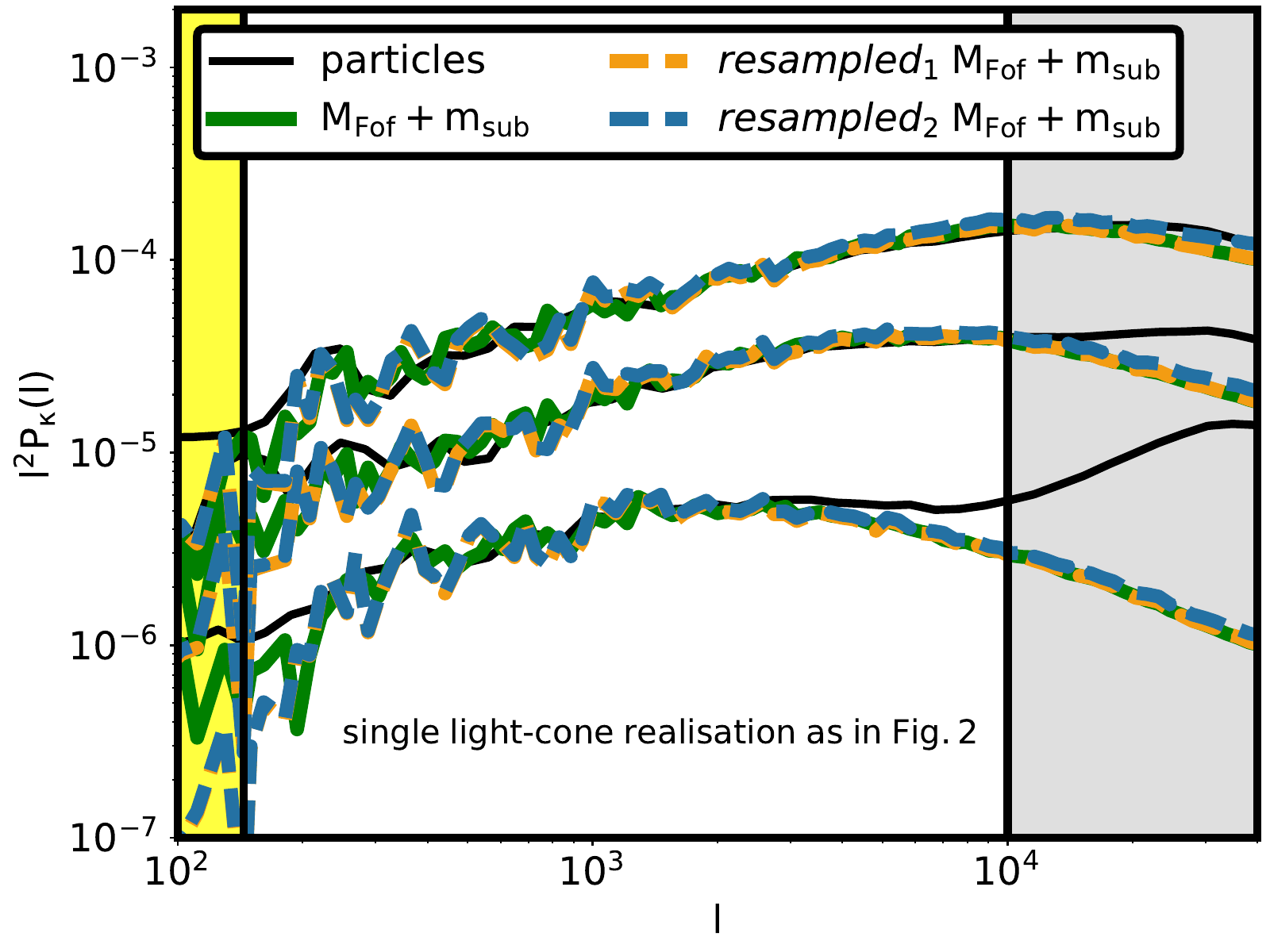}
\caption{{Convergence power spectra for sources at three different
    redshifts, $z_s=0.5$, $1.4$ and $4$ from bottom to top,
    respectively. Black curves show the power spectrum of the
    convergence map computed using particles, the green ones using our
    model which includes FoF-haloes and subhaloes, the dashed orange
    and blue curves display the power spectra of the resampled maps as
    discussed in the text.}\label{lastfpower}}
\end{figure}

In Figure~\ref{lastfpower} we display the power spectra of the
resampled maps normal and lognormal as discussed above in the text,
the orange and the blue curves display the two cases,
respectively. From the figure we can notice that since the power
spectrum is small compared to unity the differences between the normal
and the one that ensures the correct power spectrum for lognormal
field is negligible.  The curves from top to bottom display the power
spectra considering sources at $z_s=4$, $z_s=1.4$ and $z_s=0.5$,
respectively.

\bibliographystyle{mn2e}

\bsp	
\bibliography{paper}
\label{lastpage}
\end{document}